\documentclass[a4paper,11pt]{article}

\usepackage{a4wide}
\usepackage[english]{babel}
 \usepackage{relsize}
\usepackage{graphicx}

\providecommand{\e}[1]{\ensuremath{\times 10^{#1}}}
\usepackage{amsmath} 
\usepackage{booktabs} 
\usepackage{caption} 

\usepackage[colorlinks,bookmarks=true]{hyperref}
\hypersetup{
    colorlinks,
    citecolor=black,
    filecolor=black,
    linkcolor=black,
    urlcolor=black
}
\usepackage{a4wide}
\usepackage{hyperref}
\usepackage{graphicx}
\usepackage{float}
\usepackage{amsmath} 

\usepackage{titlesec}

\titleformat{\paragraph}
{\normalfont\normalsize\bfseries}{\theparagraph}{1em}{}
\titlespacing*{\paragraph}
{0pt}{3.25ex plus 1ex minus .2ex}{1.5ex plus .2ex}

\title{\vspace{-5mm}NeuCoin: the First Secure, Cost-efficient\\ and Decentralized Cryptocurrency\\
  \vspace{8mm}
  \large Version 1.0\\ March 25, 2015\vspace{3mm}}

\author{
  Kourosh Davarpanah\\
  kourosh@neucoin.org
  \and
  Dan Kaufman\\
  dan@neucoin.org
  \and
  Ophelie Pubellier\\
  ophelie@neucoin.org
\vspace{3mm}}
\date{}

\begin{document}

\maketitle

\subsection*{Preface}

The NeuCoin Project hopes that this white paper will spark an honest, fact-based debate on the pros and cons of the two main systems used to provide security for peer-to-peer cryptocurrencies: proof-of-work and proof-of-stake. 
 
NeuCoin believes that its own proof-of-stake design fully solves the security and centralization problems with earlier proof-of-stake coins, as well as the mounting cost and increasing centralization problems faced by proof-of-work systems. As such, NeuCoin is the first peer-to-peer cryptocurrency, regardless of technology, that is secure, cost-efficient and decentralized in the long run.

The NeuCoin Project is inviting all members of the cryptocurrency community to make comments, suggest edits, and point out potential flaws in the paper. NeuCoin bounties will be rewarded by the NeuCoin Code foundation for all constructive input, both positive and critical (click here for details).

Note that this paper does not discuss consensus algorithms based on trusted nodes (the systems used by Ripple and Stellar) because while that protocol has similar benefits to those of proof-of-stake, it requires users to trust third parties and beyond that, depends on “trustworthy” third parties to support the network by operating nodes, which may or may not happen over time.

Also not covered in this paper is the enormous side benefit derived from not having to distribute the currency supply to proof-of-work miners to cover their operating costs. Capitalizing on this opportunity, NeuCoin distributes the currency strategically to all the participants who increase its utility and value. Please visit NeuCoin's wiki\footnote{\href{http://www.neucoin.org/en/wiki/}{neucoin.org/en/wiki}} for a summary of NeuCoin's distribution and user adoption plans.

\newpage

\begin{abstract}
NeuCoin is a decentralized peer-to-peer cryptocurrency derived from Sunny King's Peercoin, which itself was derived from Satoshi Nakamoto's Bitcoin. As with Peercoin, proof-of-stake replaces proof-of-work as NeuCoin's security model, effectively replacing the operating costs of Bitcoin miners (electricity, computers) with the capital costs of holding the currency. Proof-of-stake also avoids proof-of-work's inherent tendency towards centralization resulting from competition for coinbase rewards among miners based on lowest cost electricity and hash power. 

NeuCoin increases security relative to Peercoin and other existing proof-of-stake currencies in numerous ways, including: (1) incentivizing nodes to continuously stake coins over time through substantially higher mining rewards and lower \textit{minimum stake age}; (2) abandoning the use of coin age in the mining formula; (3) causing the \textit{stake modifier} parameter to change over time for each stake; and (4) utilizing a client that punishes nodes that attempt to mine on multiple branches with duplicate stakes.

This paper demonstrates how NeuCoin's proof-of-stake implementation addresses all commonly raised ``nothing at stake'' objections to generic proof-of-stake systems. It also reviews many of the flaws of proof-of-work designs to highlight the potential for an alternate cryptocurrency that solves these flaws.

\end{abstract}

\vspace{15mm}

\tableofcontents

\newpage

\section{Introduction}
NeuCoin's technical ambition is to create a cryptocurrency that is secure, cost-efficient and decentralized in the long run. This paper explains why NeuCoin chose proof-of-stake over proof-of-work to achieve these goals.
 

Over time, the cryptocurrency community has generally become aware of several drawbacks of Bitcoin that spring from its proof-of-work design, including:
\begin{enumerate}
\setlength{\itemsep}{0pt}
\item{the prospect of higher transaction fees in the long run in order to maintain security}
\item{the increasing centralization and corporate control of mining}
\item{the divergence of interests between miners and Bitcoin holders}
\end{enumerate}

These flaws are mounting as Bitcoin matures, and they go against Satoshi's vision of an open, decentralized network maintained by its participants. They also undermine the principal economic benefit of cryptocurrency - very low transaction fees. 


Many in the community believe that Bitcoin's network effects are strong enough to prevent an alternate cryptocurrency from achieving wide consumer adoption, even if it were technically superior to Bitcoin and solved its shortcomings. Others believe that a strong alternative to Bitcoin's consensus mechanism can be built. 

``There is more than one way in which a distributed ledger system can work, and remuneration would have to be designed in such a way as to incentivise honest participation in the system without leading to socially inefficient over-investment in transaction verification'' (\textit{Source: Bank of England} \cite[p31.]{bank})

\subsubsection*{Proof-of-stake's advantages over proof-of-work}

Proof-of-stake, the most widely considered alternate to proof-of-work, fits this bill perfectly. There are two fundamental differences between proof-of-stake and proof-of-work. 
First, in proof-of-stake, miners compete for newly issued coins based not on the amount of electricity and computing resources spent, but rather on the number of coins owned. This crucial difference effectively eliminates the operating costs incurred in proof-of-work mining, replacing them with the capital costs of holding coins.
Second, coinbase rewards (called \textit{\textit{coinstake}} rewards in proof-of-stake) are typically not a fixed amount (as in 25 per block in Bitcoin) but proportionate to the number of coins held by the miner and the period of time during which the coins used to stake have remained idle. As such, they are akin to ``interest payments'' on the miner's coin holdings.

Based on these two differences, proof-of-stake completely solves the three problems with proof-of-work that were presented above.
\begin{enumerate}
\setlength{\itemsep}{0pt}
\item{With virtually no operating costs in proof-of-stake, transaction fees can be far lower than in proof-of-work in both the short run and long run, and regardless of transaction volumes. The only reason they are above zero is to prevent transaction spam.}
\item{Proof-of-stake doesn't suffer from gradual centralization as proof-of-work does, because all proof-of-stake miners earn the same rate of return on their coins (the ``interest rate'') regardless of computing hardware or electricity costs.
}
\item{There can be no misalignment between miners and coin holders, since they are by definition one and the same.}
\end{enumerate}

Proof-of-stake also has security advantages over proof-of-work when it comes to resisting 51\% attacks from large hostile actors - those whose purpose would be to destroy the network as opposed to simply seeking financial gain. In proof-of-stake, an attacker would have to acquire 51\% of the total coin supply (the value of which would be destroyed in an attack), versus deploying 51\% of total computing power in proof-of-work (which computing power could be re-deployed elsewhere after the attack). Once a proof-of-stake coin achieves a material value, it would be extremely expensive for an attacker to buy up a large percentage of all coins.

\subsubsection*{Bitcoin supporters' rational response to proof-of-stake}

It would be economically irrational for large Bitcoin owners to be supportive of or even open-minded about an alternative cryptocurrency design that purported to solve Bitcoin's main flaws and beyond that, enabled distribution to all the participants who helped increase its utility and value. Bitcoin supporters want all demand for digital currencies to be channeled to Bitcoin only. Moreover, they trust that if enough time passes before a strong challenger arises, Bitcoin's network effects will have grown to the point where it simply won't matter if someone builds a better mousetrap.

The party line among Bitcoin supporters is that in proof-of-stake, there's ``nothing at stake''. What they mean is that since proof-of-stake mining does not consume any outside resources (electricity, computing power), miners have no costs, so nothing prevents them from endlessly trying to commit double-spends, or mining on multiple branches, no matter how low the odds of success. Since there is ``no cost'' to behaving maliciously, proof-of-stake systems are unsecure and can't even reach consensus. They ask, ``how can you have security without paying anything for it?''

What proof-of-work proponents are neglecting to see is that proof-of-stake security does have a cost: the capital cost of acquiring and holding coins. The brilliance of proof-of-stake is that it turns all coin owners into security providers, and requires any would-be attacker to purchase a large amount of the currency to attempt an attack, which would be an attack on his own wealth.

Besides ignoring the reality of capital costs, proof-of-stake critics are also prone to depicting scary-sounding attack vectors against proof-of-stake - grinding through the blockspace, rewriting history with old private keys, long-range, pre-programmed double spends - without explaining the details of how these attacks would be conducted or demonstrating mathematically that they have more than an infinitesimal chance of success. The truth is, these attack vectors do represent valid areas of concern, so it is interesting to ask why the critiques are always theoretical and never concrete.

Perhaps by leaving the critiques abstract, without even mentioning Peercoin, NXT, Bitshares, or Blackcoin, etc., with all their different parameters, the point is made that proof-of-stake's ``nothing at stake'' flaw is fundamental and nothing can be done to fix it. Or it could simply be that analyzing the odds of success of a given attack vector against a specific proof-of-stake implementation isn't worth the effort. Besides, none of the existing proof-of-stake coins have published substantial rebuttals to the various critiques. 

NeuCoin hopes that this paper - first by frankly describing proof-of-work's own serious flaws, and second, by explicitly addressing the generic critiques of proof-of-stake - will force proof-of-stake critics to engage in a real debate about the pros and cons of proof-of-work versus proof-of-stake.

\subsubsection*{NeuCoin's design answers the ``nothing at stake'' argument}

To choose the parameters and features of its own design – indeed to even pursue a proof-of-stake design in the first place – NeuCoin began by researching and mathematically modeling all the potential attack vectors against a proof-of-stake cryptocurrency: double spends, history revisions, grinding attacks, and preprogrammed attacks. In the end, NeuCoin believes that it was able to architect a proof-of-stake design that defeats all these attacks and addresses all ``nothing at stake'' issues.

There are three critical elements to its design (explained in detail in section~\ref{32})\footnote{This summary of NeuCoin's changes to existing proof-of-stake implementations uses technical terminology that will be unfamiliar to those not well versed in proof-of-stake. These terms will be explained in detail in section~\ref{31}.}:
\begin{enumerate}
\setlength{\itemsep}{0pt}
\item{\textbf{High mining incentives:} NeuCoin provides much higher mining rewards than existing proof-of-stake currencies in order to maximize the percentage of coins being mined at all times. The odds of success for all attack vectors in proof-of-stake are based on the percentage of staked coins that the attacker controls, so it is paramount to maintain a high percentage of all coins staking across time, which existing proof-of-stake coins fail to do. To further bolster mining participation, NeuCoin reduced \textit{minimum stake age} to one day (from 30 days in Peercoin) and abandons the use of coin age as a factor influencing the probability of generating a block.}
\item{\textbf{Redesigned \textit{stake modifier}:} NeuCoin chose to adapt a version of BlackCoin's \textit{stake modifiers}, which floats over time, rather than use Peercoin's design, which permanently fixes the \textit{stake modifier} after the initial \textit{modifier interval}. NeuCoin chose this design because Peercoin's design is susceptible to preprogrammed long-range attacks (described in section~\ref{334}). Moreover, the \textit{modifier interval} and \textit{selection interval} parameters were substantially adjusted to minimize the threat of grinding (described in section~\ref{333}).}
\item{\textbf{Duplicate stake punishment:} NeuCoin uses a client version developed by Michael Witrant aka ``sigmike'' (core developer of Peercoin and Technical Advisor to NeuCoin) that not only “detects” duplicate stakes so that honest nodes can reject them, but also “punishes” nodes that broadcast duplicate stakes by rejecting all blocks broadcast by the dishonest miner. }
\end{enumerate}

The majority of this paper is devoted to describing the general principles of proof-of-stake, reviewing NeuCoin's own design, and demonstrating mathematically how NeuCoin addresses the ``nothing at stake'' objections and foils all attack vectors. 
 
But before diving into that technical material, the paper will first discuss Bitcoin's proof-of-work design and its inherent problems, which are creating a clear opportunity for a competitor that can solve them.

\newpage
\section{Proof-of-work and Bitcoin}
\subsection{How proof-of-work secures the Bitcoin network}

The public has a grand view of how Bitcoin - the digital currency with no banks or governments backing it - is secured. Bitcoin ``miners'' are envisioned to be computer experts racing against each other to solve complex math problems using cryptographic algorithms on powerful computers. In the public's imagination, the security is driven by the complexity of the math problems, the cryptography, the network's computing power and the army of computer expert miners.

The digital currency community knows that the reality of Bitcoin mining is more prosaic. Systems administrators connect computers to the internet and run a software program that races to perform a function as fast as it can, the only purpose of which is to generate a pseudo-random number. The ``complex math problem'' is in fact just the pseudo-random number generator purposely designed to use a lot of computing resources. The owner of the first computer to generate a number below a certain threshold - a ``proof of work'' - gets to add a block to the \textit{\textit{block chain}}\footnote{The \textit{block chain} is the public ledger of all Bitcoin transactions that have ever happened.} and is rewarded with free Bitcoin. 

The ``proof-of-work'' in Bitcoin mining is actually just ``proof'' that a miner did the ``work'' of running the software program and using electricity and computing power. It is proof that the miner incurred costs and spent money. Consensus security in Bitcoin is based not on the complexity of the math problems or the advanced cryptography, but on the amount of electricity and computing power spent by the miners.\footnote{Bitcoin and Mooncoin (a cryptocurrency hoped to be used on the moon) both use the same technology and design (since Mooncoin is a fork of Bitcoin); the only difference is that there is very little hash power securing Mooncoin.} The more resources spent, the more secure the network. Cutting the spending cuts the security. \textbf{Cost equals security.}

This may not sound so impressive to the public and mainstream media, but in fact it was the first practical - and brilliant - solution to a very difficult problem. To help explain Bitcoin's proof-of-work solution we will quote from a paper by Bitcoin researcher Andrew Poelstra\footnote{We quote from Poelstra's paper entitled ``Distributed Consensus from Proof of Stake is Impossible'' - the proof-of-work community's favorite critique of proof-of-stake - which we rebut in section~\ref{3}.}. 

Securing a peer-to-peer cryptocurrency requires that nodes of a distributed asynchronous network (the size of which is unknown and the nodes of which are anonymous) reach a consensus on the time ordering of messages.

As Poelstra writes\cite{distributedconsensus}:

``The reason that this consensus is needed is called the double-spending problem. That is, in any decentralized digital currency scheme there is the possibility that a spender might send the same money to two different people, and both spends would appear to be valid. Recipients therefore need a way to be assured that there are no conflicts, or that if there are conflicts, that the network will recognize their version as the correct one. A distributed consensus on transaction ordering achieves this: in the case of conflict, everyone agrees that the transaction which came first is valid while all others are not.''

In the case of Bitcoin, Poelstra continues:

``It can be mathematically proven that given only an asynchronous network it is impossible to achieve distributed consensus in a cryptographically guaranteed way\cite{Fischer}. Bitcoin achieves the impossible by weakening its requirement from cryptographic guarantee to a mere economic one. That is, it introduces an opportunity cost from outside of the system (expenditure on computing time and energy) and provides rewards within the system, but only if consensus on an unbroken transaction history is maintained.''
``To accomplish this, Bitcoin provides a way to prove, for each candidate history, (a) that opportunity cost was forfeited, and (b) how much. This is a so-called proof-of-work. Furthermore, the work proven includes that of all participants who worked on the history. The consensus history is the one with most total work (at least as far as it has propagated through the network — our weak synchronicity requirement means that the consensus on the most recent part of the history is uncertain). Since the consensus history is the only one containing spendable rewards for work done, this means (a) that provers have an incentive to work on the same history that other provers are, and (b) individual provers can't take control of the history because they need their peers' contribution.''

The technical specifics of creating a ``proof-of-work" are actually quite simple. A miner must find a tuple that satisfies the following equation:
$$ hash(blockheader,nonce)\leq target$$

The $blockheader$ is a data structure that contains information about the block, which itself encapsulates information about the Bitcoin ledger and transactions. The $target$ is a value shared by the network, adjusted so that on average new blocks are mined every ten minutes. The $nonce$ is an integer. 

The only way for a miner to add a block is to be the first one to find a hash below the target. The only way to do that is to try his luck as many times as possible by incrementing the nonce and checking the newly computed hash. The rate at which a miner generates blocks is proportional to the computing power he deploys relative to the rest of the network. 

Bitcoin's proof-of-work system provides great security so long as no hostile actor controls 51\% or more of the aggregate computing power in the network. If any entity did control a majority of the network's hash power, it could alter transaction history, or effectively shut the network down by refusing to record new transactions. 

Since there are no barriers to entry in mining, the system reaches an equilibrium where mining expenses are roughly equal to mining revenues. That is, miners continue to add incremental hash power so long as the expected revenues are higher than the incremental expenses (where ``expected'' revenues are the product of the coinbase reward and the probability of earning it).

Bitcoin enjoys high levels of security today because the high price of Bitcoin incentivizes miners to deploy a huge amount of hash power\footnote{The hash rate deployed on March 12th on the Bitcoin network was 391,367,666 GH/s or 391 PH/s \textit{source:} \url{http://www.blockchain.info/charts/hash-rate}), ie more than 5,000 higher than the combined power of the top 500 supercomputers that exist in the world (\textit{source:} \url{http://www.top500.org/statistics/list/})}. In effect, security in Bitcoin and all proof-of-work designs is 100\% dependent on and correlated to the costs incurred by miners generating their proofs of work. 
In the next section, we consider the implications of a system where its security is a function of its operating expenses. 

\subsection{Problems with proof-of-work}
\subsubsection{Costs of security and transactions in Bitcoin}

It is commonly believed that the principal economic benefit of Bitcoin and digital currencies in general is their cost advantage over legacy payment systems and their potential to offer lower cost transactions.

As Marc Andreessen put it in his widely read article \href{http://dealbook.nytimes.com/2014/01/21/why-Bitcoin-matters/}{Why Bitcoin Matters}: 

``Put value in, transfer it, the recipient gets value out, no authorization required, and in many cases, no fees. That last part is enormously important. Bitcoin is the first Internet-wide payment system where transactions either happen with no fees or very low fees (down to fractions of pennies). Existing payment systems charge fees of about 2 to 3 percent – and that's in the developed world. In lots of other places, there either are no modern payment systems or the rates are significantly higher.''

But does Bitcoin really enable extremely low-cost transactions? When one contemplates hundreds of thousands of computers racing against each other to perform billions of pseudo-random computations per second - ``proofs of work'' - in hopes of generating a low enough number to be rewarded free Bitcoin, the value of which is equal to the entire network's costs … it is hard to imagine that this is a cost-efficient system.

\begin{figure}[H]
\centering
\includegraphics[width=96mm,height=63mm]{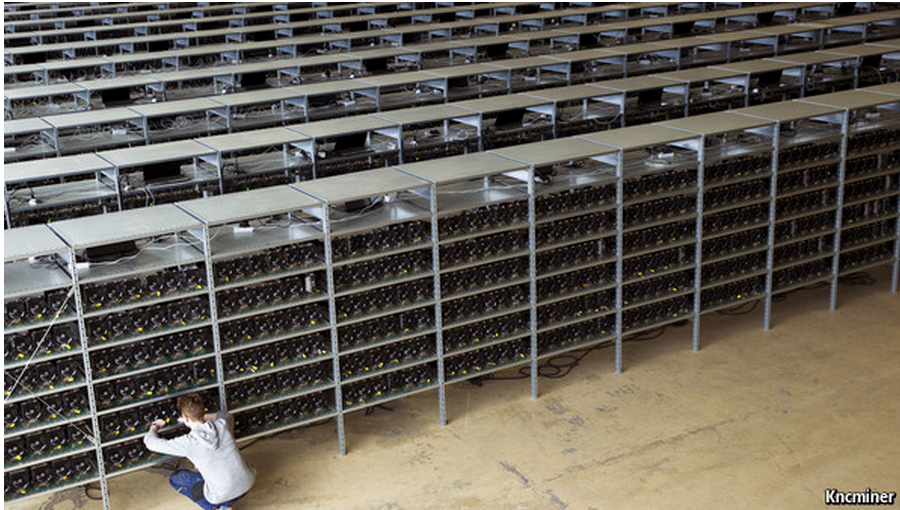}
\caption{A technician tending a warehouse full of mining rigs dedicated to 
running a hashing function that outputs a pseudo-random 256-digit number.}
\end{figure}

As discussed above, Bitcoin's network security is based on the total hash rate\footnote{The hash rate is a unit of measurement corresponding to the number of calculations per second a computer can perform. The hash rate of the Bitcoin network equals the combined hash rates of all computers mining Bitcoin.} of all the miners in the network. And the aggregate hash rate deployed by miners is strictly a function of financial payments made to miners. Hence, network security is directly determined by payments to miners. For security to stay high, payments to miners must stay high.

Miner payments are actually composed of two components: coinbase rewards (the new Bitcoin that is minted every block and rewarded to the miner who creates the block) and transaction fees. From Bitcoin's inception in January 2009 through November 2012, coinbase rewards equalled 50 Bitcoin per block. In November 2012, they were cut to 25 per block, which will continue until roughly July 2016, when they will be cut in half again to 12.5 per block. This halving continues every 4 years indefinitely, so that coinbase rewards will decline to 0.2 Bitcoin in 25 years and to practically zero in 50 years.

Here is the simple formula for computing the value of miner payments each day:
$$ \text{\textit{daily payment}} = \text{\textit{coinbase reward}} + \text{\textit{transaction fees}} $$
$$= [\text{\textit{coinbase per block}}]\times [\text{\textit{blocks per day}}]\times [\text{\textit{Bitcoin price}}]$$ 
$$+ [\text{\textit{number of transactions per day}}]\times[\text{\textit{Average transaction fee}}] $$

Using actual numbers from March 6th 2015\footnote{\textit{source:} \url{http://www.blockchain.info}} and an average \textit{block time} of 10 minutes: coinbase reward of $25 \times 144 \times \$271$ plus transaction fees of $96,611 \times \$.04$, which equals: $\$975,600 + \$3,864 = \$979,464 $

Today fully 99.6\% of the payments to miners are composed of newly created Bitcoin and just 0.4\% is in the form of transaction fees. The actual cost per Bitcoin transaction today is \$979,464/96,611, or \$10.14. 

Yes, those hundreds of thousands of mining rigs performing pseudo-random computations to create proofs of work end up costing a lot for every transaction they process.

Bitcoin end users today are shielded from (and oblivious to) these high costs because the high \textit{coinbase} rewards today pay for the costs of the system. But what will happen over time as the \textit{coinbase} rewards keep getting cut in half? Where will the miner payments come from that provide Bitcoin's security? Looking back at the formula, security will decrease unless:
\begin{enumerate}
\setlength{\itemsep}{0pt}
\item{the price of Bitcoin rises}
\item{the number of transactions rises}
\item{the price of transaction fees rise}
\end{enumerate}

Bitcoin supporters maintain that both the price of Bitcoin and transaction volumes will multiply 100- or even 1,000-fold from present levels to keep transaction fees low and security high in the long run. Looking at last year's numbers, those assumptions seem rather optimistic. 

\begin{figure}[H]
\centering
\includegraphics[width=110mm,height=63mm]{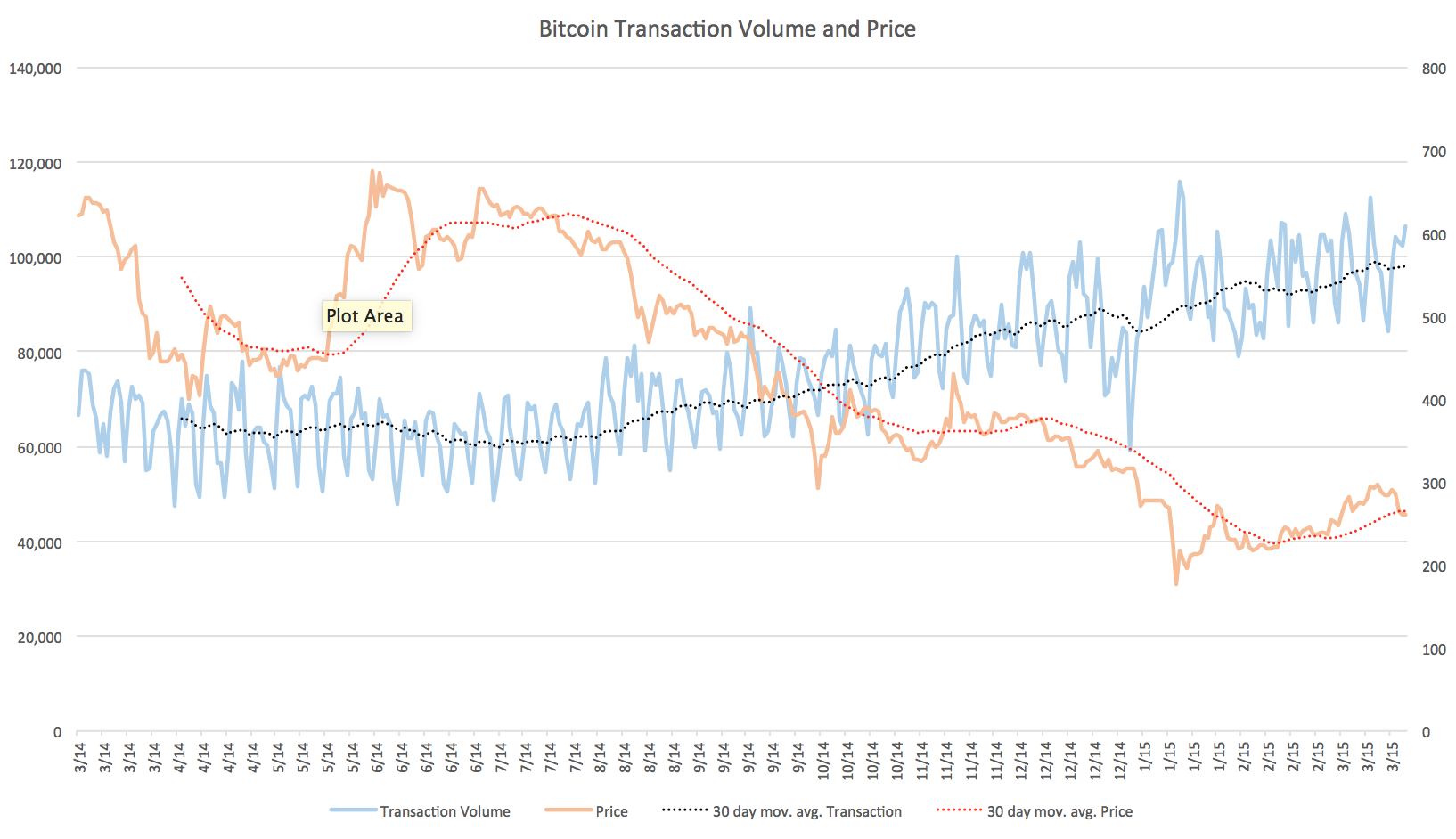}
\caption{Bitcoin transaction volume and price from March 2014 to March 2015 (\textit{source: blockchain.info}).}
\end{figure}

Meanwhile, even if user demand for transactions were to soar, there is the problem that Bitcoin's current block size doesn't allow transaction volumes of more than 5-10X today's levels. As Gavin Andresen, the Bitcoin Foundation's Chief Scientist (and Bitcoin's original core developer) put it:

``Once we get to one megabyte we've got to make blocks bigger. If we don't, transaction fees will just rise and rise and rise to a point where only rich people can afford to transact on the Bitcoin network... How is going to be the tricky bit. It's going to be hard to get consensus on exactly how far we should raise the block size.''\footnote{Gavin Andresen's keynote at Bitcoin 2014 in Amsterdam in May 2014}

Getting consensus with Bitcoin miners who have veto power over changing the block size (or making any other change to the Bitcoin protocol requiring a hard fork)  will indeed by ``tricky''. Miners may prefer smaller block sizes and higher transaction fees. Moreover, in its six years of existence, Bitcoin has not been able to achieve the necessary consensus to actually implement a hard fork a single time. 

In conclusion, it is very hard to agree with Marc Andreessen that Bitcoin enables ``no or very low transaction fees (fractions of pennies)''\footnote{Meanwhile, Marc Andreessen's venture capital firm is the largest investor in Ripple, a Bitcoin competitor with a \$1.1 billion market capitalization which does not use proof-of-work mining and thus can offer near zero transaction fees in the long run.}. Secure, low cost transactions are incompatible with a system whose security is a direct function of its cost.

This could be disappointing to some, but maybe the greatest benefit of Satoshi's vision - of a decentralized network maintained by its participants without third party financial intermediaries - is not low costs but rather the decentralization itself.

\subsubsection{Increasing centralization}

In Bitcoin's early years, mining was highly decentralized among thousands of individuals using consumer-grade computers - and this true peer-to-peer nature was considered one of Bitcoin's core benefits: anyone was able to participate in making the \textit{block chain} secure, removing the need for a central authority to validate and guarantee transactions. This was Satoshi's original vision. But as Bitcoins came to have a material value, Bitcoin mining has turned into a highly competitive business, with no barriers to entry, huge economies of scale, and only one dimension along which to compete: cost.

As a result, slowly but surely from 2010 through mid-2013, but after that in a landslide, hobbyist and small-scale miners have gotten knocked out of business. It's a very simple dynamic. When a miner with access to capital enjoys lower overall costs than the competition, he simply buys more computer power, driving up the difficulty of earning mining rewards, and knocking the least efficient competitors out of business. Lower costs can either take the form of lower cost hash rate or lower cost electricity. Bitcoin mining is rapidly becoming controlled by a handful of companies with tens of millions of dollars of the most efficient \textit{ASICs}\footnote{\textit{Application-specific integrated circuit.} Mining ASICs are specifically designed to compute hashes and have made mining with a personal computer obsolete.}, operating from facilities with the very lowest electricity and cooling costs on the planet (e.g. Iceland and northern Sweden). 

Centralization of proof-of-work mining is problematic for three reasons: 

First and foremost, decentralization is one of the primary tenets of Bitcoin: a prime reason that Satoshi created it and its most attractive feature to many of its holders. Decentralization is what enables the removal of pesky, intrusive and costly middlemen. As Bitcoin's increasing centralization becomes clearer and clearer to the crypto-community, many of its supporters may abandon it in favor of a more decentralized solution.

Second, centralization of mining represents a severe security risk. Any entity (or entities working together) that controls 51\% or more of the network's computing power can seriously harm the network. As Gavin Andresen put it: ``One of the things a 51\% attacker can do is prevent any transactions or new blocks from anybody besides themselves from being accepted, effectively stopping all payments and shutting down the network''\cite{gavin51}.

Third, when there are only a few, highly-capitalized entities that control the network, the entire network becomes susceptible to government control through regulation of these few entities. 
Governments often allow individuals to freely carry out certain activities among themselves but strictly regulate businesses that facilitate those same activities. As such, while governments might take a hands off approach to tens of thousands of geographically dispersed miners processing transactions for peers, they could decide to heavily regulate the few giant miners that dominate transaction processing.  

The centralization of Bitcoin mining was not what Satoshi or many Bitcoin supporters wanted, but it's a fact. As a consolation, at least one would expect those powerful miners that control the \textit{block chain} to do what's best for Bitcoin. Let's explore that.

\subsubsection{Divergence of interests between miners and coin holders}

In the early days, when mining \textbf{was} decentralized, miners and Bitcoin owners were mostly the same people, so miner interests and Bitcoin holder interests were basically the same. 

Over time though, Bitcoin miners have become a distinct group from Bitcoin holders. Most miners sell the Bitcoins they earn right away in order to pay for energy costs, to recover their investment in hardware before their mining equipment becomes obsolete, and to invest in more advanced next generation \textit{ASIC} chips.

The dominant miners today are large companies run by corporate executives backed by institutional investors looking to take them public. As such, they should be expected to behave just like other large companies: dedicated to maximizing their profits and minimizing their legal and regulatory risks, and looking at developing additional revenue streams, such as opportunities to sell their data or offering different pricing tiers for faster transactions. 

As of today, the interests of corporate miners and Bitcoin holders are still relatively well aligned, because both groups have the same primary, overriding goal - to increase the price of Bitcoin. Miners share this goal with Bitcoin holders because today, 99.6\% of miner revenues are in the form of newly created Bitcoin.

However, as one looks into the future, as miner revenues shift away from \textit{coinbase} rewards to transaction fees, miners and holders will no longer share the same primary goal. Bitcoin holders will still want to maximize the value of Bitcoin. Miners will want to maximize their revenues from transaction fees and other sources.

Transaction fees are just one issue. As mentioned earlier, miners also have veto power over any changes to Bitcoin's protocol. It is hard to foresee how the dynamics will play out between Bitcoin's miners, core developers and holders, but given miner control over transaction fees and protocol changes, it's safe to assume that the interests of the large corporations that control Bitcoin mining will take precedence.

\subsubsection{Summary of Bitcoin's proof-of-work problems}

Bitcoin is not cost-efficient, it is becoming more and more centralized, and the miners who control the network have diverging interests from Bitcoin holders. 

As Vitalik Buterin, inventor of Ethereum, put it in December 2014:
``What Bitcoin is doing is paying \$600M/year [...] on a 5 of 10 multisig. Ultimately there are maybe 5 or 10 [...] mining companies that control the entire network. [...] It's this incredibly inefficient protocol where miners are literally competing to see who can waste the most resources the fastest and on the other hand it's not getting us much decentralization.''\footnote{https://www.youtube.com/watch?v=qPsCGvXyrP4 at 12:50}

The preceding section on Bitcoin's and proof-of-work's problems was not meant to argue that Bitcoin is doomed, or even that it will not remain the leading cryptocurrency for the foreseeable future. Bitcoin was the pioneer and enjoys substantial network effects stemming from its large community of investors, developers and companies that are building out its ecosystem and creating meaningful value propositions for users.

However, Bitcoin's serious and growing problems should be kept in mind when evaluating the viability of alternate cryptocurrency designs. In particular, these real problems should be weighed against the hunch-based arguments and folklore that have been used to claim that other designs are unviable or even outright ``impossible'' because they are ``not secure enough'', or ``not decentralized enough''. With that in mind, let's now review the main alternative to proof-of-work: proof-of-stake.

\section{Proof-of-stake}\
\label{3}

\textit{
Note: This section uses Peercoin\cite{peercoin} as its primary reference, rather than other leading proof-of-stake implementations such as NXT or Bitshares, for two reasons: 
\begin{itemize}
\item{Peercoin was the first proof-of-stake cryptocurrency to launch and publish source code (2012, versus 2014 for NXT and Bitshares). It has received the most discussion and attention and has been forked the most times}
\item{Peercoin was the proof-of-stake implementation that NeuCoin chose to fork, for the reasons above and because the NeuCoin team felt most confident in using Peercoin as the basis for its own secure proof-of-stake design.\footnote{This does not mean that NeuCoin does not believe that NXT and/or Bitshares could also be viable implementations to achieve adequate security in a proof-of-stake design, just that NeuCoin determined that Peercoin was the safest, most direct path.}}
\end{itemize}
}

\subsubsection*{Introduction to proof-of-stake}

In proof-of-work, the task of reaching consensus on transaction history is delegated from coin holders to miners, who spend computing power and electricity, but need not own any coins. In proof-of-stake systems, the task of reaching consensus is not delegated to third parties; it is done by the coin holders (the ``stake'' holders) themselves.

In proof-of-stake, mining nodes generate blocks in proportion to their stake in (i.e. ownership of) the currency itself. Every second, every \textit{stake} has a certain probability to generate a block and this probability is proportional to its size. The hashing operation is done over a limited search space (specifically one hash per unspent transaction output, or \textit{UTXO}, per second) instead of an unlimited search space as in proof-of-work. 

Instead of providing security based on \textbf{operating costs} as in proof-of-work, proof-of-stake provides security based on the \textbf{capital costs} of investing in the currency.\footnote{Some Bitcoin proponents insist that the ownership of currency cannot be considered a ``cost''. They should consider the opportunity cost of converting fiat money (which can earn interest) into digital currency and the risk one takes of devaluation while he holds the digital currency. If converting fiat currency into digital currency were ``costless'', people should be willing to do so even without the promise of gains.}

This one critical change from proof-of-work is what allows extremely low transaction fees in the long run - without relying on enormous growth in transaction volume or the price of tokens. 
 
The second major difference between proof-of-stake and proof-of-work mining is the formula for computing mining rewards. Instead of using a fixed reward (e.g. 25 Bitcoin per block), proof-of-stake miners generally earn a rate of return on the \textbf{number of coins} being staked and the \textbf{amount of time} they have been idle\footnote{Time ``idle'' means the amount of time that has elapsed since the staked coins last earned a \textit{coinstake} reward or were sent from one address to another.} (akin to an interest rate). 

Acting together, these two changes allow a proof-of-stake cryptocurrency to remain decentralized in the long run. Miners have no potential to gain advantages over each other by investing in powerful, specialized computers or by accessing lower cost electricity. In addition, a miner who owns 10\% of the currency supply will not earn appreciably higher returns over time than a miner who owns 0.01\% of the currency supply - they both earn the same interest rate even if the larger coin holder receives payments more frequently.\footnote{The larger coin holder has only one small advantage: he receives more compounded interest, which is insignificant in comparison to the overall \textit{coinstake} rewards.}

\subsection{How Peercoin's proof-of-stake works in contrast to proof-of-work}
\label{31}

In proof-of-work coins like Bitcoin, miners use computing power to extensively modify the input of the hash. By design, the nonce creates a quasi-infinite search space to find a block header that produces a valid proof of work.\footnote{Even without the nonce, which is used in practice to ``grind'' through a large number of block headers to find a suitable hash, a miner could very well obtain the same result by modifying a number of other parameters.}

In Peercoin, a miner's chances of creating a block are based on his stake alone regardless of computing power. This is accomplished by limiting miners to the rate of one hash per second per stake.

Thus, to create a valid ``proof of stake'' and generate a block, a miner must:
\begin{itemize}
\item{Control coins (technically, an unspent transaction output, or \textit{UTXO}) that have been idle for a minimum time period (the \textit{minimum stake age})}
\item{Find a \textit{kernel} (the equivalent of the block header in proof-of-work) that satisfies Peercoin's mining formula\footnote{In Peercoin's mining formula, $\text{\textit{balance of \textit{UTXO}}} \times \text{\textit{time weight}}$ is sometimes written as $coins \times age$, or as just $coinage$. In Peercoin, \textit{time weight} is assigned a value of 0 for 30 days following the time that stake received a \textit{coinstake} reward or was moved from one address to another, and then grows from a value of 1 to 60 over the following 60 days, where it is capped.}:}
\end{itemize}

$$ hash(\textit{kernel})\leq target \times \text{\textit{balance of \textit{UTXO}}} \times \text{\textit{time weight}}$$

\vspace{5mm}
In the mining formula, the $target$ is the same for the entire network and is readjusted at every block. The $balance$ of the \textit{UTXO} is the size of the stake. The \textit{time weight} relates to the amount of time since the coins were touched. Note that in contrast to proof-of-work, where all miners are hoping their hashes fall below a given shared target, proof-of-stake miners are hoping their hashes fall below a value that is unique to each stake (the larger the stake, the easier to create the proof). 

The \textit{\textit{kernel}} is a data structure composed of the following parameters:
\begin{itemize}
\setlength{\itemsep}{0pt}
\item{\textit{nTimeTx}: current timestamp (incremented every second)}
\item{\textit{nTxPrevTime}: Timestamp of the \textit{UTXO}}
\item{\textit{nPrevoutNum}: Output number of the \textit{UTXO}}
\item{\textit{nTxPrevOffset}: Offset of the \textit{UTXO} inside the block}
\item{\textit{nTimeBlockFrom}: Timestamp of the block which provided the \textit{UTXO}}
\item{\textit{nStakeModifier}: a 64-bit string seeded from the \textit{block chain}\footnote{The generation mechanism of the \textit{stake modifier} and its role in proof-of-stake mining are described in details in the following paragraphs.}}
\end{itemize}

The current timestamp (\textit{nTimeTx}) modifies the \textit{kernel} each second to provide one chance per second per \textit{UTXO} to generate a block. 

So that every stake performs differently, the \textit{kernel} needs to include information specific to the stake. The timestamp of the \textit{UTXO} (\textit{nTxPrevTime}), its output number (\textit{nPrevoutNum}) and its offset in the block which provided its first confirmation (nTxPrevOffset) fulfill that role. The reason why all three of them are included is simply to reduce the chance of nodes having the same \textit{kernel} (which could result in them generating blocks at the same time).

If all miners could be trusted to behave honestly, the above parameters would be sufficient. However, such a design would have three serious vulnerabilities:
\begin{enumerate}
\item{First, miners could mine on multiple chains to avoid the chance of having their blocks \textit{orphaned}, which would undermine the ability of the network to reach consensus.}
\item{Second, it would allow users to pre-compute future proofs-of-stake by calculating hashes using future \textit{timestamps}. This could encourage miners to only mine their stakes at the time they were likely to earn a \textit{\textit{coinstake}} reward, severely curtailing the mining participation rate over time. It could also allow miners to attack the network by colluding.}
\item{Third and most problematically, users could gain advantage over other miners by using computational power to ``grind'' through \textit{\textit{kernels}} to identify virtual \textit{UTXOs} that would mine very quickly or at some specific time in the future. If it were profitable to gain advantage using computational power, then the system would degrade into a proof-of-work situation, with all of its shortcomings.}
\end{enumerate}

To correct the first issue, Peercoin built a feature into its client called the \textit{duplicate stake detection mechanism} that detects when miners attempt to mine on multiple chains and rejects their blocks mined with duplicate stakes.\footnote{Proof-of-stake critics raise the concern that there could be a “tragedy of the commons” caused by large numbers of selfish miners colluding and altering their clients to both transmit and accept duplicate stakes, preventing the network from reaching consensus. The NeuCoin team does not share this concern. This sort of selfish behavior could only yield negligible gains but would destroy the currency if it became widespread. Regardless, NeuCoin's modification to Peercoin's client completely addresses the concern of nodes mining on multiple chains by penalizing those that attempt to do so. See section~\ref{32} for further discussion.}

Peercoin attempted to address the second and third issues at the protocol level by including a parameter which included the timestamp of the block which provided the $\textit{UTXO}$ (\textit{nTimeBlockFrom}). It also included the current proof-of-stake \textit{difficulty} from which the network's target is derived. These features were intended to obfuscate the future performance of the stake, but did not perform as well as necessary.

Recognizing that Peercoin was still vulnerable to the precomputation of future proofs and grinding attacks, seven months after Peercoin's launch, creator Sunny King introduced a final parameter, called the \textit{stake modifier} (\textit{nStakeModifier}). The next two sub-sections explain this complex parameter, which is widely misunderstood in the digital currency community at large and even among proof-of-stake's own proponents and detractors.  

\subsubsection*{Peercoin's \textit{stake modifier} parameter}

In brief, the \textit{stake modifiers} for a given \textit{UTXO} is calculated using bits from blocks that are generated \textbf{after} the miner receives the \textit{UTXO}. Furthermore, each \textit{stake modifiers} modifies a large group of stakes - all of them that have their first confirmation in a six-hour window in the case of Peercoin. 

Given this design, a miner cannot precompute a future proof when he first receives a \textit{UTXO} because he does not yet have the \textit{stake modifiers}, one of the required parameters of the \textit{\textit{kernel}}.\footnote{In the case of Peercoin, the \textit{stake modifier} for a stake can be calculated 9 days after receiving the stake. At that point in time, a user can compute when in the future the stake is likely to generate a proof-of-stake.}.

As for how the \textit{stake modifiers} impacts attempts at grinding through the blockspace, there are two variants to consider: 
\begin{enumerate}
\setlength{\itemsep}{0pt}
\item{\textbf{Dishonest miners working on the main chain.} The \textit{stake modifiers} completely thwarts this attack. Recall that without \textit{stake modifiers}, the miner could grind through various virtual \textit{UTXOs} to find a \textit{kernel} that performs well with a timestamp in the near future. With the advent of \textit{stake modifiers}, given that (a) \textit{kernels} cannot be hashed without \textit{stake modifiers}, and (b) that \textit{stake modifiers} cannot be computed for some period after a miner receives a \textit{UTXO}, there is no way to grind through virtual \textit{UTXOs} to find a high-performing \textit{\textit{kernel}}.}
\item{\textbf{Dishonest miners working on their own forks.} This case, in which miners have complete control over the block space on their own fork, represents the most virulent form of the ``grinding attack'' referred to by proof-of-stake critics. The threat is that attackers could grind through different \textit{\textit{stake modifiers}} to try to find those that enable the attacker to perform better than the rest of the network. A detailed discussion of this attack, along with how NeuCoin's design addresses it, is presented later in section~\ref{333}. For now it will be noted that the \textit{stake modifiers} was designed with the express intent of making this attack impractical by limiting the room to maneuver that an attacker has even operating on his own branch. More specifically, the \textit{stake modifiers}:
\begin{enumerate}
\setlength{\itemsep}{0pt}
\item{is the only parameter through which an attacker could grind;}
\item{cannot be predicted;}
\item{modifies a large group of stakes at once, so that attackers cannot grind through stakes one at a time. He is stuck having to grind through \textit{\textit{stake modifiers}} in hopes of finding one that enables a large group of stakes to perform better than the network as a whole. This will be shown to be near impossible unless the attacker owns a near-majority of all staked coins.}
\end{enumerate}
}
\end{enumerate}

\subsubsection*{Mechanics of \textit{stake modifier} generation in Peercoin}

\textit{NOTE: The material in this sub-section can be skipped by non-technical readers. It is necessary for following the later discussions regarding how NeuCoin redesigns Peercoin's \textit{stake modifier} to defeat preprogrammed attacks and counter the grinding attack.}

The mechanics of generating the \textit{stake modifier} parameter are complex. They are explained below in order to provide a basis for understanding what would be involved in a grinding attack, in which an attacker grinds through \textit{stake modifiers} in an attempt to find one that enables a large number of his stakes to outperform all the stakes in the rest of the network.

In the case of Peercoin, the \textit{stake modifier} is a 64-bit number that can be computed 9 days after the \textit{UTXO} is first received. Hence, there are $2^{64}$ (approximately $10^{19}$) possible \textit{stake modifiers} that an attacker could grind through.

A new \textit{stake modifier} is computed every \textit{modifier interval} (\textit{nModifierInterval}), which is 6 hours in Peercoin. The \textit{stake modifier} is computed for the first block of this 6-hour long window, and all the other blocks within this window inherit the same \textit{stake modifier}. Since Peercoin's \textit{block time} is 10 minutes, there are approximately 36 blocks that share each \textit{stake modifier}. All of the stakes first confirmed within any of these 36 blocks inherit the same \textit{stake modifier}.

The 64 bits in the \textit{stake modifier} are generated by first selecting 64 blocks from a time window (\textit{nSelectionInterval}) that begins 9 days before the beginning of the \textit{modifier interval} (at the time the miner receives the \textit{UTXO}) and ends when the \textit{modifier interval} begins. The selection of the blocks happens over 64 incrementally longer windows contained in the \textit{selection interval}. One block is selected from each time window. 

The graph below depicts the 64 subdivisions of the \textit{selection interval}.

\begin{figure}[H]
\centering
\includegraphics[width=110mm,height=63mm]{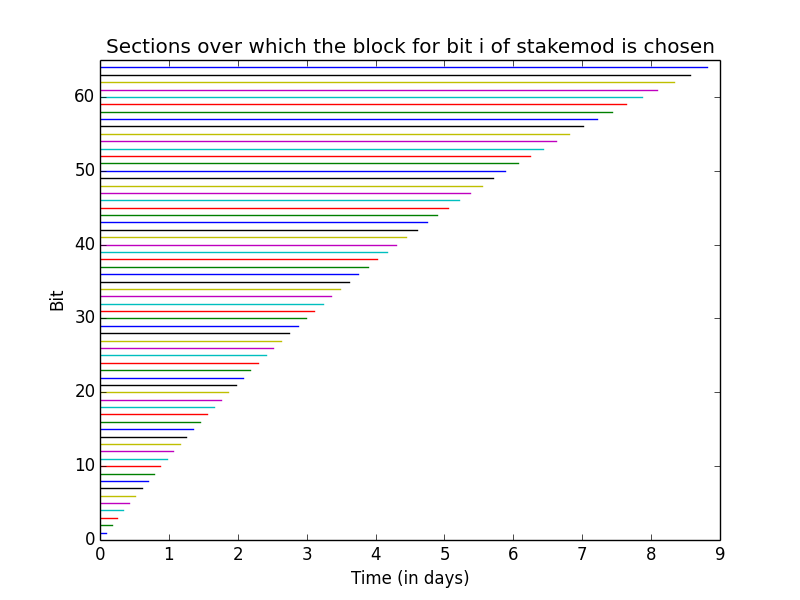}
\caption{Subdivisions of the \textit{stake modifier} \textit{selection interval} over which the block contributing to the $i$-th bit is chosen.}
\end{figure}

\newpage
Each of the 64 selected blocks contributes one bit to the \textit{stake modifiers} as described below.

The first bit is computed as follows:
\begin{itemize}
\setlength{\itemsep}{0pt}
\item{Within the shortest time window in the chart above, for each of the $\sim12$ blocks therein:}
\begin{itemize}
\setlength{\itemsep}{0pt}
\item{The hash of the \textit{kernel} used to create the block (the \textit{hashProof}) and the previous \textit{stake modifiers} are concatenated}
\item{The resulting string is then hashed}
\end{itemize}
\item{All the ($\sim12$) resulting hashes are compared}
\item{The block which provided the lowest hash is selected and the least significant bit of this hash is selected}
\end{itemize}

To compute the second bit:
\begin{itemize}
\setlength{\itemsep}{0pt}
\item{The second shortest time window is considered}
\item{The previously selected block is ignored}
\item{We go through the same steps we went through for the first bit. This time the block is selected from among $\sim24$ blocks }
\end{itemize}

This process is repeated across 64 time windows until the 64 bits have been generated. 

The complexity of the generation mechanism was required in order to create a strong interdependency between the parameters of the \textit{UTXO} used in the \textit{kernel} and the \textit{stake modifiers} as well as between the consecutive \textit{\textit{stake modifiers}}.

\subsubsection*{Mining rewards and transaction fees}

The entire technical discussion of Peercoin above related to how miners create ``proofs of stake'' in order to be able to generate blocks. Here we briefly consider the financial rewards of mining. 

As mentioned earlier, miners in Peercoin earn rewards (\textit{coinstakes}) that are akin to interest rates. The precise mining reward formula is as follows:
$$ \textit{coinstake} = \text{\textit{balance of \textit{UTXO}}}\times \frac{ \text{\textit{days idle}} }{365}\times 1\% $$

\textit{Balance of \textit{UTXO}} is the number of coins, \textit{days idle} is the amount of time since the coins last earned a \textit{coinstake} reward or were transferred from one address to another (technically, it equals the current timestamp minus the timestamp of the \textit{UTXO}). 

There are two interesting things to note. First, miners do not earn any transaction fees, which are equal to $.01$ Peercoin per transaction. Rather than paying these fees to miners, Peercoin elected to actually destroy them. Its rationale was to partially offset the growth of the currency supply due to \textit{coinstake} rewards. Peercoin's creator Sunny King put a very high priority on achieving a long term inflation rate close to zero.

But the much bigger point about Peercoin's reward formula is that miners only earn a 1\% annual rate of return on their staked coins. As we will see later, this low rate is a serious problem.

\subsubsection*{Is Peercoin's proof-of-stake design secure?}

In other words, does Peercoin answer the ``nothing at stake'' critiques? Does it both (a) prevent attackers from altering history and (b) maintain consensus on the order of transactions on a single \textit{block chain}?

Before answering this question, let's recall the context. Bitcoin itself is not cryptographically secure\cite{Fischer}. It only tries to be ``economically secure'' in the sense that an attacker would have to spend more than he could gain through an attack. Furthermore, it is fully acknowledged that any actor controlling 51\% of Bitcoin network's hash power could effectively shut Bitcoin down. In addition, Bitcoin has some flaws - inherently high operating costs, increasing centralization, and diverging interests between miners and coin holders - which Peercoin would correct - so perhaps some loss of security in Peercoin relative to Bitcoin would be an acceptable tradeoff. 

Even with all of these considerations, our conclusion would still be that Peercoin's design is only somewhat secure - and significantly less secure than that of Bitcoin. The rationale for our conclusion will be explained in the following section~\ref{32}, where we review all of the changes that NeuCoin has made to Peercoin's design. In our opinion, the factor that detracts the most from Peercoin's security is that its design fails to incentivize a large proportion of coins to be staked at any one time. It also suffers from issues related to its use of coin age in block generation. Lastly, it fails to foil attempts to precompute proofs of stake due to the static nature of its \textit{stake modifier} design. 

Whether or not Peercoin's creator Sunny King would agree with our specific conclusions is unknown. It is likely, however, that he shares the view that Peercoin's design is not currently ``secure enough''. This can be inferred from the fact that three years after its creation, Peercoin still uses centrally broadcast checkpoints that prevent any possible changes to the part of \textit{block chain} earlier than the checkpoint. The checkpointing is done several times per day by way of Sunny King digitally signing the \textit{block chain}.

The great thing about \textit{checkpointing} is that even proof-of-stake's fiercest critics will concede that it does secure the currency against attacks. The downside of checkpointing is that it is a form of centralization. It is not the case that the entity that performs the checkpointing has the power to control the \textit{block chain}, but it is the case that all users of the currency are relying on this ``trusted entity'' to provide a necessary layer of security. 

Unfortunately for Peercoin, checkpointing has severely hurt market acceptance of the currency partly because the checkpointer is the anonymous creator Sunny King himself. There are obvious and near-fatal problems with trusting an anonymous person with a necessary security function. What happens if that anonymous person dies or becomes incapacitated, sells off his or her stake in the currency, or simply loses interest? 

\newpage
\subsection{NeuCoin's proof-of-stake design}
\label{32}

NeuCoin believes that its own proof-of-stake design is fully secure - and will not need to use checkpoints. This section describes all of the principal changes that NeuCoin made to Peercoin's design and explains how they increase security and address Peercoin's vulnerabilities. Following this section, the paper will review all of the attack vectors against proof-of-stake designs and demonstrate how NeuCoin's design stands up to these attacks.

NeuCoin significantly modified six aspects of Peercoin: 

\begin{enumerate}
\setlength{\itemsep}{0pt}
\item{\textbf{Mining reward rates:} NeuCoin dramatically increased \textit{coinstake} rewards for mining in order to maximize the percentage of coins being mined at all times, which is the bedrock of security in any proof-of-stake cryptocurrency. NeuCoin's rewards start at a 100\% annual interest rate and decline steadily over a 10-year period to a 6\% rate - versus just 1\% per year in Peercoin.}
\item{\textbf{Minimum stake age:} NeuCoin's design uses a 1.6 day \textit{minimum stake age} versus 30 days in Peercoin. This also has the effect of increasing mining participation.}
\item{\textbf{Role of coin age in the mining equation:} NeuCoin does not utilize coin age (\textit{dayweight}) in the mining equation as a factor for determining the probability of generating a block. This change also increases mining participation.}
\item{\textbf{Block time}: NeuCoin uses a \textit{block time} of 1 minute, versus 10 minutes in Peercoin, which improves user experience and enhances security against some attack vectors.}
\item{\textbf{Stake modifier:} NeuCoin chose to adapt BlackCoin's \textit{stake modifier}, which floats over time, rather than Peercoin's, which permanently fixes the \textit{stake modifier} after the initial stake interval for a given set of \textit{UTXOs}. NeuCoin chose this design because it believes Peercoin's design is susceptible to preprogrammed long-range attacks (described in section~\ref{334}). Modifier interval and \textit{selection interval} were substantially adjusted relative to both BlackCoin and Peercoin in order to reduce the effectiveness of grinding through \textit{stake modifiers}.}
\item{\textbf{Duplicate stake punishment:} NeuCoin uses a client version developed by Michael Witrant, aka ``sigmike'' (core developer of Peercoin and Technical Advisor to NeuCoin), that not only “detects” duplicate stakes so that honest nodes can reject them, but also “punishes” nodes that broadcast duplicate stakes by rejecting all blocks broadcast by the dishonest miner. This revision completely addresses the concern that proof-of-stake designs cannot reach consensus due to miners mining on multiple forks.}
\end{enumerate}

\subsubsection*{Mining reward rates: 100\% annual rate declining to 6\% per year}

The key to high security in all proof-of-stake coins is maintaining a high percentage of coins staking over time. Therefore, it is critical to provide high compensation to miners for staking their coins.

One severe weakness of Peercoin is that it suffers from extremely low mining participation, with typically less than 10\% of the currency supply staked at any one time. This means that an attack requiring control of 51\% of the staked coins only requires control of 5.1\% of the total currency supply. There are several factors that contribute to the low mining rates in Peercoin, but the principal one is miniscule rewards for mining: just a 1\% annual return. The small amount of mining that does take place in Peercoin is most likely driven more by selfless efforts to support the currency than by the financial rewards of mining.\footnote{This issue of low mining incentives is even worse for NXT than for Peercoin. NXT has no \textit{coinstake} rewards at all. Instead, the only financial reward for mining is to earn transaction fees, which amount to just \$26 worth per day at NXT's market price on March 17, 2015.}

In contrast, NeuCoin provides very high compensation to miners, especially during the early years following its launch. Specifically, NeuCoin's \textit{coinstake} rewards begin at a 100\% annual rate and decline in a linear fashion over ten years after which they reach a 6\% annual rate, where they remain indefinitely. These high reward rates will incentivize a large number of miners to stake a large number of coins. In addition, these high reward rates create incentives for cloud mining operators to provide services to NeuCoin holders not interested in mining themselves, where both the mining operator and the NeuCoin holder can receive substantial rewards.

\begin{figure}[H]
\centering
\includegraphics[width=110mm,height=63mm]{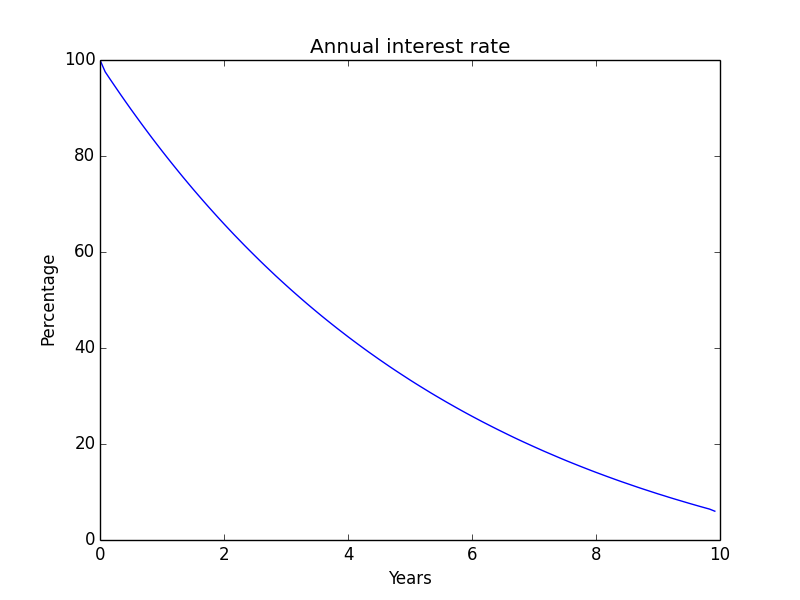}
\caption{Evolution of the annual interest rate during the first 10 years.}
\end{figure}

Some observers might consider NeuCoin's high reward rates to cause unacceptably high inflation. In fact, NeuCoin utilizes high inflation deliberately in its early years to compensate users for holding and staking the currency while its value, utility and security are lowest and its risk is highest.

Given the mining reward rate of 6\% from year 10 onwards, and assuming a 50\% coin staking rate, long-term currency supply growth is projected to be 3\% per year. Given that this is slower than fiat currency monetary growth (which averages the rate of economic growth plus the rate of inflation), in the long run there will be a mild deflationary effect relative to fiat money. This rate of inflation is highly preferable to outright deflation for any digital currency that aspires to serve as an actual unit of account and means of exchange for commerce as opposed to merely a store of value.

\subsubsection*{Minimum stake age: 1.6 days}

On top of high mining rewards, the second important modification NeuCoin made to increase mining participation was to reduce \textit{minimum stake age} to 1.6 days, versus 30 days in Peercoin. This change significantly decreases the number of coins that are ineligible for mining because they have done so recently (more recently than the \textit{minimum stake age}). It also reduces the barrier to entry associated with having to wait for such a long time before being eligible to start mining with any given stake.

It would have been preferable to have a \textit{minimum stake age} of even less than one day - one hour for instance. For the same reasons listed just above, an even lower \textit{minimum stake age} would have tended to increase the mining participation rate even further. Unfortunately, at a certain point, lower \textit{minimum stake ages} can begin to increase the risk of a grinding attack (discussed in section~\ref{333}). NeuCoin believes that 1.6 days represents the best tradeoff.\footnote{In fact, one of NeuCoin's top research priorities is finding ways to prevent grinding attacks with lower mininum stake ages, which would allow NeuCoin to reduce \textit{minimum stake age} to 12 hours or less.}

\subsubsection*{Coin age: zero effect}

NeuCoin does not utilize coin age (\textit{day weight}) in the mining equation as a factor for determining the probability of generating a block. NeuCoin's mining equation is simply:$$ hash(\textit{kernel}) < target \times \text{\textit{balance of \textit{UTXO}}} $$

versus Peercoin's mining equation: $ hash(\textit{kernel}) < target \times \text{\textit{balance of \textit{UTXO}}} \times \text{\textit{time weight}}$
\vspace{1mm}

In Peercoin, \textit{time weight} was the parameter that reflected coin age, equalling 0 up to the \textit{minimum stake age} (30 days) and then growing from a value of 1 to 60 over days 31 to 91, at which level it is capped.

NeuCoin's deletion of the \textit{time weight} parameter brings two significant benefits for security but at the cost of a loss of decentralization. Let's consider the benefits first.

First, \textit{time weight}'s deletion removes the opportunity for attackers to \textit{supercharge} their stakes. An attacker can keep his node offline until he reaches the maximum \textit{time weight} of 60 (after waiting 91 days from receiving the stake) and then carry out an attack with far more block generating power than he would normally have. This \textit{supercharging} effect is especially potent if an attacker were to use it in combination with a preprogrammed, long-range attack, described in detail in section~\ref{334}.

\textit{Time weight}'s deletion also tends to increase mining participation. Allowing stakes to accumulate \textit{time weight} for block generation encourages miners to keep their nodes offline and come back online after accumulating significant \textit{time weight} to very quickly collect the reward. This behavior is compounded with Peercoin's 30 day \textit{minimum stake age}, which requires nodes to wait 30 days before earning a \textit{coinstake} reward. But at 31 days, a miner's odds of earning a reward are very low because he is competing with nodes that have high \textit{time weight}. Rational (albeit selfish) behavior would be to wait 90 days before mining, at which time the reward will likely be earned quickly. Obviously, the percentage of coins being staked over time suffers greatly if even dedicated miners find it advantageous to leave their nodes offline the majority of the time.

As mentioned above, there was also a downside to disregarding coin age in the mining equation - a loss of decentralization. 

In fact, using coin age in the mining equation is a centerpiece of Peercoin's design. The benefit of the time-weight parameter is that it significantly reduces the amount of time it takes \textbf{very small coin holders} to mine blocks. This goal was clearly a top priority for Peercoin. Recall Peercoin's mining equation, where time-weight = 0 for the first 30 days and then grows from values of 1 through 60 over the following 60 days. Now consider two holders of Peercoin, one who has owned 1,000 Peercoins for 90 days while the other has owned 50,000 Peercoins for 31 days. Due to time-weight's role in Peercoin's mining equation, the small coin holder actually has a higher chance of mining than the large coin holder with only 1/50th of his stake. This benefit was undoubtedly a major selling point for Peercoin when it was introduced to the cryptocurrency community August 2012, which then consisted almost entirely of proof-of-work miners. Many of these miners must have become frustrated by the fact that it could take years for a small miner to earn a coinbase reward in Bitcoin.
All that said, it must be recalled that all miners in NeuCoin do still benefit from coin age in the \textit{coinstake} reward equation, which is the same as it is in Peercoin, (except for the fact that in NeuCoin the interest rate starts at 100\% and declines to 6\%, versus a flat 1\% in Peercoin):

$$ \textit{coinstake} = \text{\textit{balance of \textit{UTXO}}} \times \frac{\text{\textit{days idle}}}{365} \times x\% $$

Consider two NeuCoin miners with the same number of coins (balance of \textit{UTXO}). Miner A is lucky and receives one \textit{coinstake} after 45 days ($\text{\textit{days idle}} = 45$) and a second \textit{coinstake} 15 days later (days idle = 15). Miner B receives only a single \textit{coinstake} after 60 days ($\text{\textit{days idle}} = 60$). How much better off is Miner A over Miner B? Miner B's single \textit{coinstake} was \textbf{almost} equal to the sum of the two \textit{coinstakes} received by Miner A, but not quite. The reason for the discrepancy is that Miner A earned a tiny bit of incremental compounded interest: 15 days of interest on the first \textit{coinstake} reward he received on day 45.\footnote{To illustrate how small a factor this is, consider the \textbf{effective} annual interest rate of a 10\% simple annual interest rate compounded at different intervals: compounded quarterly: 10.38\%; compounded monthly: 10.47\%; compounded weekly: 10.51\%; compounded daily: 10.52\%} In addition to this slightly higher amount of total \textit{coinstake} rewards, Miner A also had a small benefit in terms of liquidity. That is, Miner A could have sold his coins after 45 days without sacrificing any accumulated coin age (equivalent to accrued interest). Miner B had to wait an extra 15 days.

In sum, in abandoning time-weight, NeuCoin chose the benefit of higher security at a cost of creating small advantages for large miners over smaller miners, which would tend to lessen decentralization. However, the final thing that should be noted is that only small miners will experience this disadvantage to a material degree. Given NeuCoin's parameters for \textit{minimum stake age} and \textit{block time}, a miner owning 1/50,000 of the total currency supply would be expected to earn a \textit{coinstake} once per month.\footnote{Assuming $\sim85\%$ of the total number of coins mine at any given time. If the percentage were to decrease, the considered miner would earn a \textit{coinstake} more often. }

\subsubsection*{Block time: one minute}

NeuCoin chose a 1 minute \textit{block time} primarily to benefit users who appreciate one relatively quick (though not very reliable) confirmation for transaction processing. In addition, having a shorter \textit{block time} does increase security for some types of attacks.

The probability for a transaction to be reversed by an attacker depends on its number of confirmations (i.e. how deep the transaction is in the \textit{block chain}). The more confirmations, the more blocks the attacker needs to create on his competing fork.

In the case of a simple double spend - or any attack that attempts to alter transaction history without doing anything to improve performance such as grinding or supercharging stakes - shorter \textit{block times} reduce the attacker's odds. In these cases the attacker is hoping to get lucky over a short period of time thus he is hurt by having to create more blocks in a given time period.

However, in the case of a grinding attack or a deep reorganization, where the attacker has to build a very long chain, his only hope is to be able to consistently create blocks faster than the main chain (because luck won't help over a long time horizon). In these cases, \textit{block time} is irrelevant.

The reason that NeuCoin did not choose an even shorter \textit{block time} is that the \textit{block chain} fork rate (the probability that a miner's block is orphaned) increases when \textit{block time} decreases. With a 10 minute \textit{block time}, the fork rate is approximately 1.9\%. With 1 minute, it is closer to 17\%. Going down to 10 seconds would cause a fork rate of roughly 70\%\cite{decker2013information}. Reaching consensus in these conditions would be difficult and inefficient, as there would be near constant reorganizations. Given that getting orphaned has only limited impact on mine

Reaching consensus in these conditions would be difficult and inefficient, as there would be near constant reorganizations. Given that getting orphaned has only limited impact on miner returns in proof-of-stake, NeuCoin considered 1 minute a good tradeoff between fast transaction confirmations and \textit{block chain} fork rate.

\subsubsection*{Stake modifier: dynamic; 200 minute \textit{modifier interval}; 1.6 days \textit{selection interval}}

As described in section~\ref{31}, Peercoin's \textit{stake modifier} is generated 9 days (the \textit{selection interval}) after the first confirmation of a given stake. Although the \textit{stake modifiers} is not computable at the time the stake is received, once it is generated it does not change over time. As a result, miners can predict when they will be able to mine in the future with a given stake. This could reduce mining participation over time, but more critically, it opens the door to the preprogrammed double spend attack, described in section~\ref{334}.

BlackCoin addressed this vulnerability of Peercoin's ``static'' \textit{stake modifier} in their August 2014 hardfork. BlackCoin's protocol update changed the \textit{stake modifier}'s behaviour in two ways:
\begin{enumerate}
\setlength{\itemsep}{0pt}
\item{It caused a given stake to use different \textit{stake modifiers} over time. Rather than acquiring a static \textit{stake modifier} generated over the initial \textit{selection interval} following a stake's first confirmation, stakes utilize the \textit{stake modifier} seeded from the most recent \textit{selection interval}.\footnote{Note that this does not mean that when a stake is received, the user can already compute its \textit{stake modifier}. Indeed, since the stake is not allow to mine for \textit{minimum stake age} (which is larger than the \textit{selection interval}), none of the blocks that will be used to compute the first \textit{stake modifier} that will be included in the \textit{kernel} exist at that point.} (BlackCoin's \textit{selection interval} being 5.87 hours).}
\item{As a consequence, all the stakes being mined at any one time use the same \textit{stake modifier}. The \textit{stake modifier} changes at every \textit{modifier interval}, 10 minutes in the case of Blackcoin. Hence, every 10 minutes a new \textit{stake modifier} is generated based on blocks from the preceding 5.87 hour window.}
\end{enumerate}

Put simply, BlackCoin's update caused the \textit{stake modifier} to become dynamic, to shift over time. This change made it impossible to precompute when in the future a \textit{UTXO} would generate a block. 

In addition, BlackCoin's choices of values for \textit{modifier interval} and \textit{selection interval} enabled BlackCoin to have a much lower \textit{minimum stake age} than PeerCoin. Having a lower \textit{minimum stake age} tends to cause a higher mining participation rate.

Unfortunately for Blackcoin, however, their choice of a 10 minute \textit{modifier interval} made them more susceptible to a grinding attack (reviewed in section~\ref{333}) than need be the case. The potential benefits an attacker can gain by grinding through \textit{stake modifiers} depends on the number of blocks contained in the \textit{modifier interval}. In essence, during this period, the attacker must be able to find more blocks than the rest of the network. With BlackCoin's 10 minute \textit{modifier interval}, the attacker must grind through \textit{stake modifiers} until he finds one that enables him to generate ten blocks faster than the rest of the network. NeuCoin believes that this is not an adequate defense against a grinding attack (discussed in section~\ref{333}).

NeuCoin chose a \textit{modifier interval} of 200 minutes and a \textit{selection interval} of 1 day. As will be shown in section~\ref{333}, these parameter choices would require an attacker using grinding to acquire at least ~31\% of the staked coin supply. Choosing a \textit{modifier interval} of 400 minutes would have increased the required coins for a grinding attack up to ~36\%. 800 minutes would have raised the requirement up to 40\%. The reason NeuCoin chose 200 minutes and not more is that longer \textit{modifier intervals} require longer \textit{minimum stake age}. And as discussed above, longer \textit{minimum stake ages} lead to lower mining participation rates.

NeuCoin is continuing to research the inter-relationship between \textit{modifier interval} and \textit{minimum stake age}. Its goal is to find improvements to its design that would further diminish the advantage that an attacker could gain from grinding, ideally to the point where the attacker would need the same 51\% of the staked coins that he would need in any other attack vector. An additional defense mechanism NeuCoin is considering is to modify the client so that it does not accept reorganizations deeper than $h$ hours worth of blocks. Since an attacker necessarily accumulates a significant lag when creating a fork, this would prevent the possibility of any malicious reorganization of the transaction history.\footnote{The downside of this defense is that new nodes would have to rely on trusted nodes from whom they download the \textit{block chain}. The NeuCoin Code Foundation, which is controlled by NeuCoin holders, would at least be an excellent entity to fulfill this role.}

\subsubsection*{Punitive duplicate stake detection mechanism}

The main branch of ``nothing at stake'' critiques against proof-of-stake relates to a lack of security against an attacker seeking to alter transaction history. The second branch, which is more subtle, argues that proof-of-stake systems will not reliably be able to maintain consensus on transaction history. This critique contends that since miners bear no operating costs, there is nothing to prevent them from mining on multiple chains. They would do so in order to avoid the possibility of their blocks getting orphaned. If this behavior were to become widespread, it could create a situation in which many competing forks of similar length coexist, making it impossible to identify a single fork as the main chain.

To address this concern, Peercoin built a feature into its client called the ``duplicate stake detection mechanism'' that detects and rejects blocks mined with duplicate stakes, which works as follows:
\begin{itemize}
\item{When a node receives a block, it checks to ensure that the proof used to create the block has not also been used to create another block previously received by the node (a ``duplicate'' stake). If the node detects the duplicate use of a stake, it simply discards the second block received\footnote{ Note that the reason nodes are able to detect users mining on two chains is because the proofs used to mine on both chains are the same. This is a consequence of the design of the \textit{kernel} which doesn't include the hash of the previous block or the root of the Merkle tree. }.}
\item{All nodes that use the unmodified client act this way. Therefore, a block created with a duplicate stake will not propagate through the network.}
\end{itemize}

Proof-of-stake critics make the point that since this mechanism is implemented only at the client level, nothing stops greedy nodes from running modified clients that would allow mining with duplicate stakes on all possible forks. Even though honest nodes would reject their blocks built with duplicate stakes, if other greedy nodes also modified their clients, the greedy nodes could collude by accepting each other's blocks built with duplicate stakes. 

NeuCoin does not consider this scenario at all realistic. The potential gains from acting dishonestly are too miniscule, and the only way the miniscule gains are earned is when this antisocial behavior becomes widespread, which would seriously diminish the value of the currency.\footnote{Given that proof-of-stake rewards are variable (akin to an interest rate), the only possible gain from mining on a secondary chain would be the compound interest earned on one \textit{coinstake} reward to the next that had been lost due to a block getting orphaned. This de minimus return would only be earned in the unlikely event that a block mined on a duplicate stake were propagated to another dishonest miner who created a legitimate block on top of the illegitimate one. Critics should consider the behavior of Bitcoin miners, who could increase their chances of creating blocks that do not get orphaned by not including any transactions at all in their blocks. A 1mb block has a $\sim13\%$ higher chance of getting orphaned than an empty one due to network lag issues\cite{gavintx}. To compensate, the block's transaction fees should equal at least $25\times13\%\sim3.2BTC$. In reality, the number is approximately $\sim 0.5 BTC$. Yet miners do not mine empty blocks because doing so would hurt the whole network and value of Bitcoin. }

Nevertheless, NeuCoin solves this theoretical problem completely by using a client that \textbf{penalizes} nodes that attempt to mine using duplicate stakes. Michael Witrant (aka ``sigmike''), core developer of Peercoin and Technical Advisor to NeuCoin, has adapted Peercoin's duplicate stake detection mechanism and turned it into a \textbf{punitive} duplicate stake detection mechanism. Quite simply, when a node receives a block generated with a duplicate proof-of-stake, it discards not only the second block but also the first one received using the duplicate stake. Therefore, whenever a miner tries mining on multiple chains, all of his blocks get rejected by honest nodes. In this system, dishonestly mining on multiple chains yields no possibility of greater rewards (rather the guarantee of lower rewards).

\subsection{How NeuCoin's design prevents attacks on transaction history}

The proof-of-work community has dismissed proof-of-stake because its miners don't have operating costs, so they have ``nothing at stake''. Given this root issue, proof-of-stake critics have described four ways that attackers could alter transaction history: simple double spends, history revisions, grinding attacks and preprogrammed attacks. Each of these attack vectors are summarized briefly below. Following the summaries, the remainder of the paper will probe each of these attack vectors in detail: explaining how they work against proof-of-stake designs in general, describing how NeuCoin's own design stands up to them, and demonstrating their odds of success with mathematical models.

The various ways that miners could try to modify transaction history in proof-of-stake can be broken down into four categories:
\begin{enumerate}
\item{\textbf{Simple double spend}

The widely and casually held view in the proof-of-work community - that it is ``costless'' to attacking a proof-of-stake currency, so attackers will keep trying until they succeed - would suggest that proof-of-stake designs are vulnerable to simple double spend attacks. Proof-of-stake critics with deep knowledge of the design might quietly concede that this simple attack wouldn't work, but they have allowed the proof-of-work community to believe otherwise. Therefore, the double spend attack will be explained in depth first to correct this widely held misconception, and second, in order to demonstrate the basic mechanics of an attack on transaction history in proof-of-stake, which remain the same in more sophisticated attacks. }

\item{\textbf{History revision using old private keys}

A more serious critique concerns the possibility of rewriting transaction history using ``old private keys'' (technically, using stakes that are no longer unspent at present time, or in layman's terms: using coins that are no longer owned). Moreover, the old stakes could have been owned by a third party (who since sold them). The disconcerting upshot is that an attacker never needs to own the currency at all (so it really is ``costless''). This section will show that when creating a fork at an earlier point in transaction history with old private keys, the attacker will be starting the attack too many blocks behind to ever be able to catch up.  }
\item{\textbf{Grinding attack}

The grinding attack - where the attacker uses computational power to grind through \textit{kernels} in hopes of discovering ones that enables him to outperform the main chain - is the most serious threat to proof-of-stake coins. This paper has already discussed how Peercoin attempts to defeat this threat through the \textit{stake modifier} and also discussed the changes that NeuCoin made to Peercoin's \textit{stake modifier} parameters. The upcoming section on the grinding attack will show how NeuCoin's design makes grinding cost-prohibitive, requiring the grinding attacker to acquire a minimum of 30\% of the staked currency in order to be successful (versus requiring 51\% of the staked currency without grinding).\footnote{NeuCoin is currently researching several possible design changes that would require a grinding attacker to acquire the same minimum 51\% stake in the currency that is necessary to carry out any other attack.} }
\item{\textbf{Preprogrammed long-range attack}

This attack involves putting together a collection of stakes that will perform very well in a specific time window in the future (for instance a year or more). This is a very potent theoretical attack against Peercoin that stems from the fact that the \textit{stake modifier} of a given stake is static. NeuCoin's changes to the \textit{stake modifier} parameters completely neutralize this attack vector.}
\end{enumerate}

\subsubsection{Simple double spend}
\label{331}

A ``simple double spend'' attack consists of spending coins and recovering them after the counterparty has given a product or value in exchange for the coins. To successfully commit a double spend the attacker must be able to rewrite the transaction history. Specifically, he must be able to fork the main \textit{block chain} and extend his side branch until it is longer than the main branch which the rest of the network is working on.

A double spend attack on a proof-of-stake coin consists of the following steps, described in a way that a non-technical person can follow:
\begin{enumerate}
\setlength{\itemsep}{0pt}
\item{First of all, the attacker must own coins. A portion $X$ of these will be used for the double spend itself while the rest ($Y$) will be used to fork the \textit{block chain} and attempt to extend this side branch until it is longer than the main branch which the rest of the network is working on. }
\item{The portion of coins $Y$ must be split into a very large number of smaller stakes to maximize the attacker's chances of success. (See the subsection ``Role of \textit{minimum stake age} in attacks'' below for an explanation.)}
\item{The attacker broadcasts a transaction to the network sending $X$ coins to the merchant he plans to attack.}
\item{The merchant waits for some number of confirmations to assure himself that the transaction is valid. In Bitcoin and Peercoin, the standard practice is to wait for 6 confirmations, or approximately 60 minutes, since their blocktime is 10 minutes. We will assume that the merchant waits the same 60 minutes to be assured that the transaction is valid. Since NeuCoin's \textit{block time} is 1 minute, this equals 60 confirmations)\footnote{ For extremely high-value transactions, merchants would typically wait for several hours' worth of confirmations to increase their security.}.}
\item{While the merchant waits for his confirmations, the attacker secretly forks the main chain and begins building a side branch on the block prior to the one in which the $X$ coins sent to the merchant was first confirmed. In his side branch, the attacker includes a conflicting transaction sending the same $X$ coins back to himself.}
\item{In order for his attack to succeed, the attacker must build 61 blocks on his side branch faster than the rest of the network put together builds 60 blocks on the main chain. 
	\begin{enumerate}
	\setlength{\itemsep}{0pt}
	\item{To help conceptualize this task, let's assume that the attacker purchased 10\% of the total amount of currency being staked by the network to use for building his branch.}
	\item{Since the \textit{block time} in NeuCoin is one minute, the target is set so that the entire network - all of the stakes being mined - should on average generate one block every one minute.}
	\item{Hence, the expected time for the attacker to a generate a single block on his side branch is 10 minutes (since $Y = 10\%$).}
	\item{Meanwhile, the expected time for the rest of the network (collectively staking 90\% of the currency) to generate a block on the main chain is 1.1 minutes.}
	\item{To summarize, the attacker must build 61 blocks faster than the rest of the network builds 60 blocks, but his expected time to build \textbf{each} block is 10 minutes versus an expected time per block of 1.1 minutes for his competitor.}
	\item{Intuitively, it's clear that this is a very difficult task for the attacker, but it is hard to fathom just how difficult it is. The actual odds that the attacker will win this race are 1 chance in $\sim10^{35}$, or 1 in 100,000,000,000,000,000,000,000,000,000,000,000\footnote{The formula used to produce this result is below.} 
}
	\end{enumerate}
\item{To conclude this illustration, in the virtually impossible event that the attacker did succeed in building 61 blocks faster than the rest of the network, after the merchant sent the attacker the product, he would broadcast his side branch, invalidating the transaction he sent to the merchant and replacing it with the transaction he sent back to himself.}
}
\end{enumerate}
  
Given that this would-be attacker in the example above had to start his attack by purchasing 10\% of the total currency supply in order to have these infinitesimal chances of success, it is obvious that this attack will never happen in the real world. It should also be pointed out that it's not as though the attacker had the chance to earn a huge reward. 

His upside was simply to defraud a merchant of X NeuCoins (not the large amount Y used for the attack). And if the attacker were attempting to defraud the merchant of a large dollar value (say \$10,000 or more), the merchant would have waited for far more than one hours' worth of confirmations.

To model the double spend attack mathematically\cite{nakamoto2008bitcoin}, we consider an attacker with a portion p of the mining coins trying to attack a merchant who waits n confirmations. The probability to succeed is:

$$ P[\text{\textit{double spend}}]= 1 - \sum\limits_{i=0}^n\frac{\lambda^{i}e^{n-i}}{k!} \Big[ 1- \big(\frac{p}{q}\big)^{n-i} \Big] $$
with $p$ the portion of mining coins owned by the attacker, $q=1-p$ the portion of mining coins of the rest of the network, $\lambda=n\frac{p}{q}$ the expected number of blocks generated by the attacker (modeled as a Poisson process) assuming the rest of the network took the average expected time per block to create $n$ blocks.

\vspace{2mm}

\begin{table}[htp]
	\begin{center}
		\begin{tabular}{ccccc}
		\toprule
			{} & {1 conf.} & {10 conf.} & {60 conf.} & {120 conf.}\\
			{\textit{p}} & {(1mn)} & {(10mn)} & {(60mn)} & {(120mn)}\\
			\midrule
			1\% & $0.020$ & $1.3\e{-16}$ & $6.1\e{-95}$ & $6.8\e{-189}$\\
			5\% & $0.10$ & $1.3\e{-09}$ & $5.0\e{-53}$ & $4.4\e{-105}$\\
			10\% & $0.21$ & $1.2\e{-06}$ & $4.4\e{-35}$ & $3.5\e{-69}$\\
			20\% & $0.42$ & $0.0011$ & $1.5\e{-17}$ & $3.7\e{-34}$\\
			30\% & $0.63$ & $0.042$ & $3.7\e{-08}$ & $2.3\e{-15}$\\
			40\% & $0.83$ & $0.36$ & $0.0083$ & $0.00010$\\
			50\% & $1$ & $1$ & $1$ & $1$\\
		\bottomrule
		\end{tabular}
		\caption{Probabilities of successfully completing a double spend in NeuCoin base on the portion of staked coins used ($p$) and the number of confirmations required by the attack victim.}
	\end{center}
\end{table}

\subsubsection*{Role of \textit{minimum stake age} in attempts to revise transaction history}

Technically speaking, an attacker cannot conduct a double spend attack with a single stake because the \textit{minimum stake age} prevents a stake from mining for 1.6 days after it has generated a block. Therefore, the attacker must split his stake into:
\begin{itemize}
\setlength{\itemsep}{0pt}
\item{at least the number of confirmations used by the merchant plus one.}
\item{to increase his odds, a much greater number than the number of confirmations so his probability to mine successfully does not drop materially after mining each block.}
\end{itemize}

\newpage
Consider an attacker planning a double spend attack on a merchant who waits for 60 confirmations. Now consider three scenarios:
\begin{itemize}
\setlength{\itemsep}{0pt}
\item{Scenario A: the attacker splits his stake into 100 \textit{UTXOs}.}
\item{Scenario B: he splits his stake into 10,000 \textit{UTXOs}.}
\item{Scenario C: he splits his stake into 1,000,000 \textit{UTXOs}.}
\end{itemize}

In all scenarios, his chances of generating his first block in a given timestamp is the same. This may seem counter-intuitive. It would seem that his chances of generating a block with 10,000 \textit{UTXOs} would be higher than it would be with 100 \textit{UTXOs}. However, one must recall that the chances of mining a block are directionally proportional to the size of the stake. So in scenario A each of the 100 \textit{UTXOs} have a 100x greater chance of creating a block than the 10,000 \textit{UTXOs} in scenario B.

Now consider what the scenarios would look like after the attacker has created 20 blocks:
Scenario A: attacker has 80 \textit{UTXOs} left with which he is trying to generate 40 more blocks. Note that he has 80\% of his original stake available to mine.
Scenario B: attacker has 9,980 \textit{UTXOs} left with which he is trying to generate 40 more blocks.
Note that he has 99.8\% of his original stake available to mine.
Scenario C: attacker has 9,999,980 \textit{UTXOs} left with which he is trying to generate 40 more blocks. Note that he has 99.9998\% of his original stake available to mine.

Since the probability of success decreases with the attacker's stakes that cannot mine, it is clear that the attacker will want to split his stake into a very large number of smaller stakes in order to increase his chances. That said, we can see from our example that the gains from splitting into smaller and smaller stakes quickly diminish - improving from the use of 80\% of the original stake to 99.8\% is much more important than the further improvement to 99.9998\%.

\subsubsection{History revision using old private keys}
\label{332}

In the previous section on double spends, it was shown that even in the case where an attacker owned a significant portion of all staked coins, he still faced infinitesimal odds of success. But proof-of-stake critics have pointed out that attackers could actually attempt to rewrite transaction history without owning any of the currency. All they would need to do is acquire “old private keys.” In other words, they could use old stakes that were previously owned by third parties who had since sold them.

The history revision attack is in essence similar to  the double spend attack. In both cases, the attacker creates a fork and attempts to extend it so that it becomes longer than the main chain.

However, in a double spend attack, the attacker starts his fork with no lag. When trying to rewrite history with old private keys, the attacker necessarily has to start the attack with a long lag, specifically, at the time in the past when the old private keys controlled \textit{UTXOs}. Hence, the attacker must not simply generate $X+1$ blocks in the time it takes the rest of the network to generate X blocks, as in a double spend attack. The history revision attacker must generate $X+1+Y$ blocks, where Y is the number of blocks represented by the attacker's lag behind the main chain. In addition, in double spend attacks where the attacker is using a stake that he actually owns (say 20\% of all staked coins), the rest of the network with which he competes owns the remaining 80\%. However, when using old private keys to 20\% of the staked coins, the attacker is competing against not 80\% of the staked coins but against 100\% of them, because the attacker's old coins are now owned by new parties who mine on the main chain.

So while it is theoretically possible for an attacker to obtain access (private keys) to another person's old stake that they have already spent or sold, the downside is that the number of blocks that the attacker needs to build on his branch is daunting. Obviously, the further back in time that the attacker would begin his side branch, the closer to impossible his task becomes.

Let's consider an example and make assumptions that would give the attacker his best possible chances. Rather than going back a long time, let's just go back one hour. This is the shortest time that the attacker's accomplice (the one who gave the attacker access to the private keys to his own stake) could have possibly sold off his coins assuming that the counterparty or merchant waited for one hour's worth of confirmations.

Now let's make the assumption that the attacker's accomplice actually owned 10\% of the entire supply of the currency that was being staked. This assumption is totally unrealistic because the merchant who acquired this gigantic quantity of coins from the accomplice would have waited for far more than one hour's worth of confirmations. 

Note that these are same numbers - 10\% of the staked currency supply and 60 blocks - as in the double spend attack from the previous section. But in this attack, during the time that the attacker is generating those 60 blocks to catch up to the main chain (``revising history''), the main chain is marching forward in time generating its own blocks, which the attacker must also match and finally surpass. While in the double spend example the attacker had a one in $10^{35}$ chance of succeeding, in this history revision attack using the same 10\% of staked coins, the odds have been cut to one in $\sim10^{95}$. Clearly, going back in time to start an attack makes things \textbf{much} harder.
 
Now let's consider a scenario where the attacker gains control of old private keys for not 10\% but rather 75\% of the currency's staked coins. But we will require the attacker to start his branch four hours (rather than just one) behind the main chain. In this case, despite having access to so much old currency, the attacker's odds of ever catching up to the main chain are still less than $\sim10^{-57}$. It may seem surprising that an attacker would not be able to catch up to the main chain with anything greater than 50\% of the staked currency. However, in the case of using old private keys, the attacker with 75\% is not competing against the network with the remaining 25\% of the coins but rather against the network with 100\% (because the old private keys have new owners mining on the main chain).

For our mathematical model, let's consider an attacker wishing to perform an $n$-deep reorganisation and controlling a portion p of the coins taking part in the mining process at height $h_{max}-n$.

We can model the arrival of the blocks on the competing chain as a Poisson process. At time $t$ after the beginning of the attack, the expected number of blocks the attacker will have created is $p \frac{t}{\tau}$, while in average the number of blocks on the main chain (starting at the fork) will be $\frac{t}{\tau}+n$. Therefore, the probability that the attacker's branch is longer than the main chain at time $t$ after the beginning of the attack is:
$$P\big[ Poiss(p\frac{t}{\tau})\geq \alpha \frac{t}{\tau} +n\big]=\Gamma\big(\alpha \frac{t}{\tau} +n,p\frac{t}{\tau}\big)$$
where $\Gamma$ is the regularized incomplete gamma function, $\tau$ is the average \textit{block time} and $\alpha=1-p$ if the attacker still owns the coins used in the reorganisation attempt and $1$ otherwise.
\newpage
The following, very basic, upper bound of the cumulative probability is sufficient to show that the probability for an attacker of ever catching up with the network is infinitesimal:
$$P[Success]\leq \sum\limits_{i=1}^\infty \Gamma\big(\alpha i+n,pi)$$ 

The following graph highlights the fact that the probability to catch up with the main chain decreases exponentially with how far behind the attacker starts:
\begin{figure}[H]
\centering
\includegraphics[width=96mm,height=63mm]{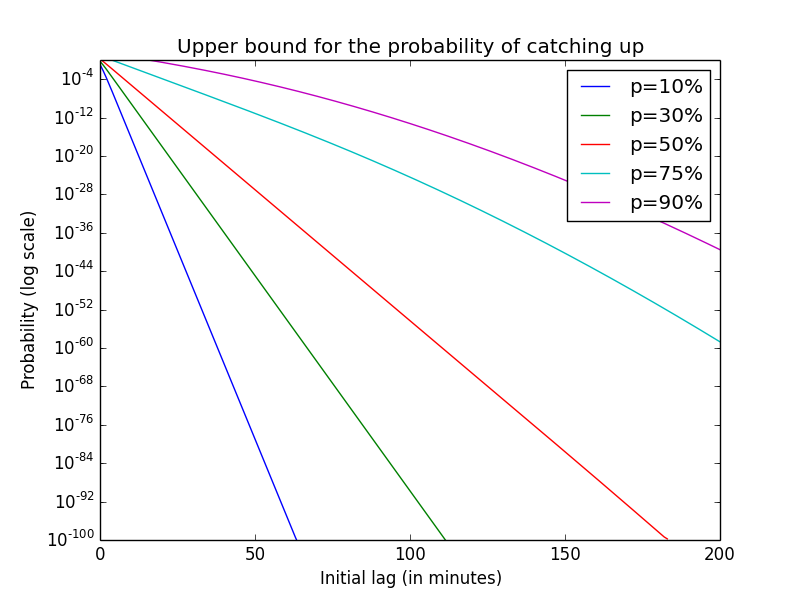}
\caption{Upper bound for the probability of catching up starting $n$ blocks behind for an attacker controlling different portions $p$ of old private keys.}
\end{figure}

The history revision attack can also be conducted with coins that the attacker still owns. In this case, the rate at which the main chain generates blocks depends on the portion of coins the attacker owns. The graph below shows that for the attacker to be successful, the attacker must own $>50\%$ of the mining coins, i.e. $>50\%$ of the mining power, like in a $51\%$ attack on Bitcoin. 

\begin{figure}[H]
\centering
\includegraphics[width=96mm,height=63mm]{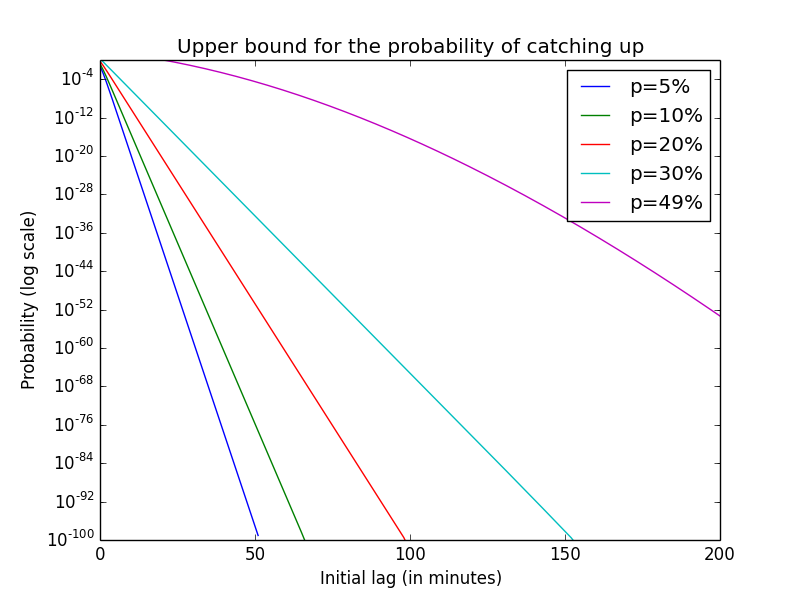}
\caption{Upper bound for the probability of catching up starting $n$ blocks behind for an attacker owning different portions $p$ of the total staked coins.}
\end{figure}

Therefore, the only way for an attacker to revise history with less than 50\% of the staked coins is to find a way to maliciously increase the mining power of his stake so that he consistently generates more blocks than the rest of the network combined. The next section describes an attack vector in which an attacker tries to gain such an advantage.

\subsubsection{Grinding attack}
\label{333}

The next idea that proof-of-stake opponents have speculated about is the possibility of gaining a significant edge against the network by ``cheaply searching the blockspace to find blocks that direct history in their favor''\cite{distributedconsensus}. More concretely, they refer to``grinding'' through \textit{kernels} in order to improve the performance of his stakes. This attack sounds very compelling to a proof-of-work miner because it is analogous to grinding through nonces as fast as possible in order to generate a proof-of-work. 
 
The rationale behind this attack is that since the inputs of the \textit{kernel} are seeded from the chain the attacker is working on, and since the attacker is working on own branch, he can shape it as he wishes and might be able to find a seed that will allow him to perform significantly better than expected.

In Bitcoin, miners ``grind'' through block headers \textbf{one by one}. Proof-of-stake designs use the \textit{stake modifier} parameter - which modifies a large number of stakes - for the precise reason of preventing attackers from being able to grind through stakes one at a time, even if the attacker is alone on his branch. The only way they can grind through stakes is indirectly, through the \textit{stake modifier}, which affects many stakes at once.

As we will see, this design feature makes the grinding attack impossible unless the attacker owns a very large portion of the total supply of coins being staked.

\subsubsection*{Step One - gaining control of the \textit{stake modifier}}

The first step of the grinding attack is for the attacker to generate blocks in order to gain influence over and eventually control the \textit{stake modifier}. This step takes time and causes
the attacker to accumulate a significant lag behind the main chain. 

The process of gaining control of the \textit{stake modifier} is very technical, difficult to express and difficult to follow. But the details do not matter. The only important takeaway is that the process takes a significant amount of time (at least hundreds of minutes).

The process is as follows:
\begin{itemize}
\setlength{\itemsep}{0pt}
\item{During the first 200 minutes\footnote{To keep things simpler we consider that the fork starts at the beginning of a new \textit{modifier interval}} (the first \textit{modifier interval}) of the attack, the attacker cannot grind through \textit{stake modifiers}.   He must use the \textit{stake modifier} that was seeded from blocks from the main chain over the preceding \textit{selection interval} (1.6 day, or $\sim2250$ minutes).}
\item{When a new \textit{stake modifier} is computed (200 minutes after the beginning of the attack), the new \textit{selection interval} will encompass the 200 minutes of the attack plus the $\sim2050$ minutes preceding the attack. Based on how many blocks the attacker managed to generate during the first \textit{modifier interval}, the attacker now has some chance of controlling a small minority of the 64 bits of the new \textit{stake modifier}. He will at most control 1 bit of the \textit{stake modifier}. (Assuming that the attacker controlled 20\% of the total staked currency, he would most likely control $\sim0.15$ bit.) Even with $1$ bit, the attacker has only gained the ability to grind through $2$ \textit{stake modifiers}. Obviously, this will not allow him to gain any advantage over the rest of the network.}
\item{The attacker repeats the same process over the next 10 \textit{modifier intervals} (2,000 minutes), after which the entire \textit{selection interval} will no longer stretch further back in time than the beginning of the attack.}
\item{The attacker is waiting to have sufficient bits under his control in order to be able to grind through a material number of \textit{stake modifiers}. His final goal is to have control over all 64 bits. This can be done in no less time than $\sim1.6$' day (the \textit{selection interval}). If the attacker owns at least $\sim3\%$ of the total staked currency, he could expect to control the \textit{stake modifier} at the end of this 1.6 day period.}
\end{itemize}

\subsubsection*{Step two - grinding through \textit{stake modifiers}}

The object of the preceding subsection was simply to demonstrate that the attacker's fork is hundreds of minutes behind the main chain. 

In determining whether a grinding attack will succeed or not, it is not material whether the attacker's chain is 400 minutes behind the main chain or 2,250 minutes behind. The only way to be able to catch up from ``far'' behind the main chain is to be able to \textbf{consistently} outperform the rest of the network. 

In the following paragraphs, we will detail the attack in order to assess what portion of the mining coins an attacker must control in order to succeed with a grinding attack. While it's obvious that an attacker controlling more than half of the mining coins would succeed, it is unclear to what extent grinding would lower this threshold. Proof-of-stake critics instinctively assume that grinding through the block space would make it possible to revise history with a minute fraction of the coins. It will be shown that a very substantial portion of the staked coin supply is necessary. 

In order to describe the grinding attack, we consider an attacker working alone on a fork and therefore having complete control over the block space. This allows him to modify the hash of blocks he creates\footnote{By incrementing the nonce for example.} in order to set the bits of a given \textit{stake modifier} as he desires.
For the sake of simplicity, we consider that the attack starts when the attacker has complete control over the current \textit{stake modifier}. 

In order to outperform the rest of the network and succeed in a grinding attack, \textbf{the attacker must be able to create more blocks than the network is expected to create during the \textit{modifier interval}.}

Here are the steps:
\begin{enumerate}
\setlength{\itemsep}{0pt}
\item{The attacker splits his stake into a very large number of \textit{UTXOs}.}
\item{The attacker will then search the \textit{stake modifier} space and for every possible \textit{stake modifier}, compute the hash of the \textit{kernel} of every one of his \textit{UTXOs}. Since the \textit{stake modifier} is a 64-bit number, there are $2^{64}$ (roughly $10^{19}$) possibilities for the attacker to grind through.\footnote{Grinding attacks require significant computing power in addition to large portions of the staked coins. For the following illustration, we assume that the attacker uses 10\% of the hash power of Bitcoin's network. Increases above this amount of hash power provide minimal gain.}}
\item{If the attacker is able to find a 64-bit string that allows him to generate more blocks than the rest of the network, he succeeds. He then proceeds to create the corresponding \textit{stake modifier}.}

\end{enumerate}

\subsubsection*{Grinding Illustration}

Consider an attacker that owns 20\% of the total supply of staked coins taking the steps listed above.

The heart of the grinding attack is that the attacker must be able to create more blocks than the network is expected to create during the \textit{modifier interval}. If an attacker owns 20\% of the staked coins, the rest of the network would be expected to generate 160 blocks (80\% times the 200 blocks in the 200 minute \textit{modifier interval}). Therefore, the attacker needs to generate 161 blocks in 200 minutes.

The attacker splits his stake into a very large number of stakes. Let's say 1 million stakes.

The attacker will then search the \textit{stake modifier} space and for every possible \textit{stake modifier} he computes the hash of the \textit{kernel} of every one of his 1 million \textit{UTXOs}. Since the \textit{stake modifier} is a 64-bit number, there are $2^{64}$ (roughly $10^{19}$) \textit{stake modifiers} multiplied by 1 million \textit{UTXOs} for the attacker to grind through.

The attacker is hoping to find a single \textit{stake modifier} for which at least 161 of his 1 million \textit{UTXOs} create a proof within the \textit{modifier interval}.

With 20\% of all staked coins, the attacker will on average find 40 blocks within this interval. The following graph shows the probability mass function for an attacker with 20\% of the coins over the \textit{modifier interval}.

With 20\% of all staked coins, the attacker will on average find 40 blocks within this interval. The following graph shows the probability mass function for an attacker with 20\% of the coins over the \textit{modifier interval}.

\begin{figure}[H]
\centering
\includegraphics[width=96mm,height=63mm]{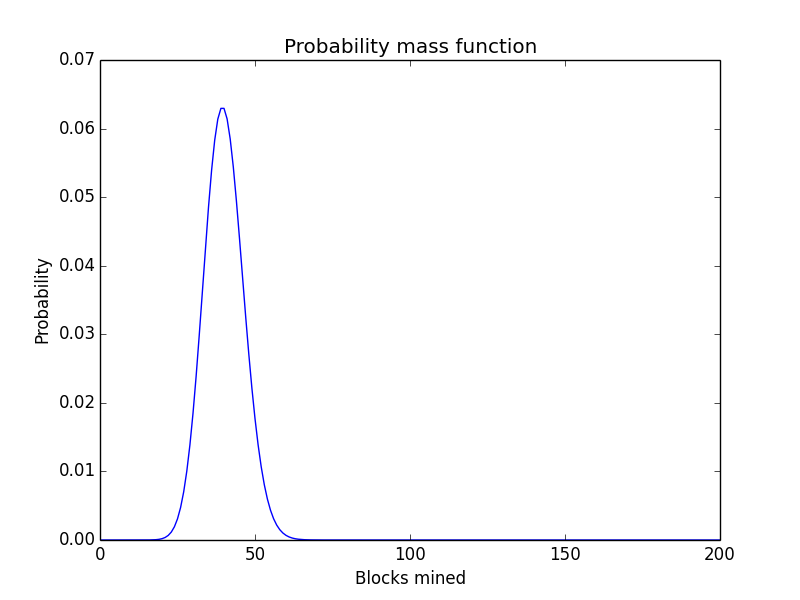}
\caption{Probability mass function for an attacker with 20\% of the mining coins over a 200 minutes time interval.}
\end{figure}

In this particular case, the probability for a given \textit{stake modifier} to create 160 \textit{kernels} that will mine is $\sim10-{46}$. 

As will be shown below, given NeuCoin's choice of \textit{modifier interval}, a grinding attack would require an attacker to own approximately 30\% of the total supply of staked currency (assuming he had access to thousands of ASICs' worth of hash power).

\subsubsection*{Mathematical model of grinding attack}

At every \textit{modifier interval}, we model the arrival of the blocks on the attacking branch as a Poisson process of intensity p times the intensity of all the mining coins combined. Therefore, the probability to create more blocks than the main chain is\footnote{We suppose that the transaction outputs used for the attack are still unspent at present time.}:
$$P(success)_{/trial}=\Gamma\Big[(1-p)\frac{T_{modifier}}{\tau},p\frac{T_{modifier}}{\tau}\Big]$$
with $p$ the portion of mining coins held by the attacker, $\Gamma$ the regularized incomplete gamma function, $T_{modifier}$ the \textit{modifier interval} and $\tau$ the \textit{block time}.

Therefore, the probability for an attacker to successfully conduct a grinding attack is:
$$P[success]=1-\bigg(1- \Gamma\Big[(1-p)\frac{T_{mod}}{\tau},p\frac{T_{mod}}{\tau}\Big]\bigg)^{min( \frac{ H_{/s}T_{mod} }{N_{stakes}},2^{64}) }$$
with $H_{/s}$ the hash rate of the attacker, $N_{stakes}$ the number of stakes the attacker controls (we need $N$ much larger than $\frac{T_{mod}}{\tau}$ but we don't take into account the fact that probability to succeed goes down if $N$ is too small), $2^{64}$ is size of the \textit{stake modifier} space.

The graph below shows the probability of successfully conducting a grinding attack as a function of the portion of coins that he controls for different amounts of hash power.
As we can see:
\begin{enumerate}
\setlength{\itemsep}{0pt}
\item{Although the grinding strategy allows to lower the cost of a 51\% attack, the attacker must still own north of $\sim30\%$ to succeed. For a coin with a significant market, this provides a much greater level of security than proof-of-work, since acquiring $\sim30\%$ of the coins would be orders of magnitude more costly than buying 51\% of the hash power of the network.}
\item{Although more hash power does decrease the percentage of coins necessary to conduct the attack, its influence is actually very limited. To draw a comparison, an attacker with $\sim51\%$ of Bitcoin's computing power would need $\sim30\%$ of the coins whereas an attacker with a single ASIC miner would need $\sim34\%$. An attacker with a hundred times the hash power of the entire Bitcoin network would need to control $\sim29\%$ of the coins.}
\end{enumerate}

\begin{figure}[H]
\centering
\includegraphics[width=96mm,height=63mm]{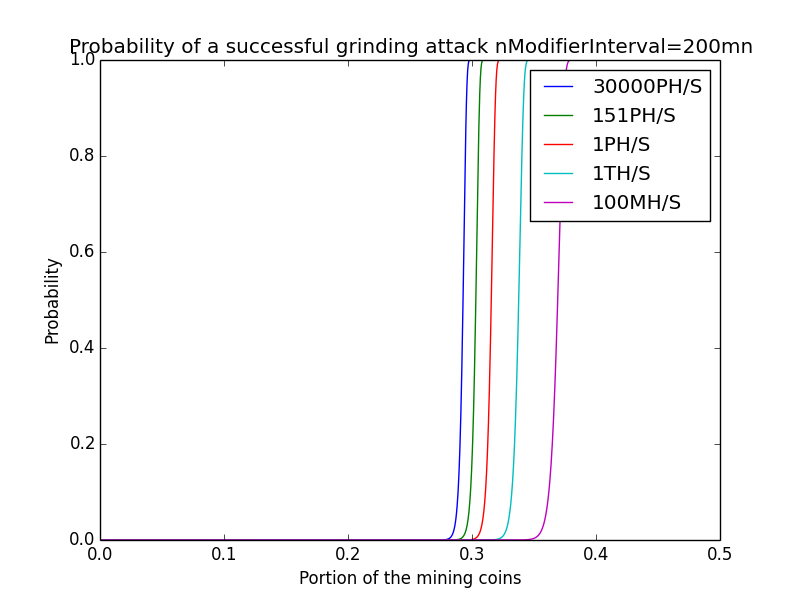}
\caption{}
\end{figure}

To conclude this discussion on grinding in proof-of-stake, here is a statistic for the critics who say it is ``costless''mini to grind through the blockspace: if an attacker owned 10\% of NeuCoin's staked currency and had access to all of the hash power of the Bitcoin network, his odds of success in a grinding attack would be…one out of $\sim10^{87}$. By the way, there are only $\sim10^{80}$ atoms in the observable universe.

\subsubsection{Preprogrammed, long-range attack}
\label{334}

The final attack vector - the preprogrammed long-range attack - has an ominous name, but is not threatening at all to NeuCoin's design.

This attack involves putting together a collection of stakes that will perform very well in a specific time window in the future (potentially a year or more). This is a very potent attack vector against Peercoin due to Peercoin's use of coin age in the mining equation (which allows ``supercharging'' stakes) and because Peercoin's \textit{stake modifier} yields a static value after the \textit{selection interval}. 

NeuCoin chose to use a dynamic rather than static \textit{stake modifier} for the principal reason of neutralizing this attack vector. With a \textit{stake modifier} that changes at every \textit{modifier interval} (200 minutes for NeuCoin), attackers cannot pre-compute future proof-of-stakes. Without any way to precompute future stakes, there is no way to put together a collection of stakes that perform well in a future time window, which is the essence of this attack.

The paragraphs below describe how this attack works against a proof-of-stake design with a static \textit{stake modifier}. 

In Peercoin, the \textit{stake modifier} of a particular stake is computable after the \textit{selection interval} and remains the same until it mines (or is used as an input in a transaction). Therefore, after the \textit{selection interval}, all the components of the \textit{kernel} are determined and the miner is able to predict at what time stamp the stake is likely able to mine. Although the target cannot be known precisely in advance, a miner can guess its future value with an acceptable margin of error.

In Peercoin, this allows an attacker to precompute future proof-of-stakes in order to carry out a long-range attack. The steps to conduct such an attack are as follows:
\begin{enumerate}
\setlength{\itemsep}{0pt}
\item{The attacker splits his coins into a large number of stakes and chooses a distant time window (e.g. 1 year or more in the future) during which he wants to conduct the attack.}
\item{After $T_{SM}$, the \textit{stake modifier} is generated and the attacker can compute the hashes of all the \textit{kernels} for time stamps included in the considered time frame.}
\item{The attacker keeps all the stakes that have a high probability of mining within that time frame and then resends all the remaining stakes back to himself (in order to modify their \textit{kernels})}
\item{He repeats the two previous steps repeatedly, in each cycle retaining the stakes that will perform well during the attack window, until the time of the targeted attack window.}
\item{Once the attack window is reached, the attacker will be able to create proofs and generate blocks with the stakes he kept. If he has a high enough number of stakes that can generate blocks (say more than 60 to fool a merchant waiting for 60 confirmations), he may be able to perpetrate an attack.}
\end{enumerate}

\newpage

\section{Conclusion}
This paper has shown that NeuCoin's carefully constructed proof-of-stake design, derived from Sunny King's Peercoin, which itself was derived from Satoshi Nakamoto's Bitcoin, is secure, cost-efficient and decentralized in the long run.
 
The paper has also demonstrated several drawbacks of Bitcoin's proof-of-work design – including higher transaction fees in the long term, increasing centralization of mining and a divergence of interests between miners and Bitcoin holders – and shown how proof-of-stake technology completely addresses these drawbacks by introducing two crucial differences:
\begin{enumerate}
\setlength{\itemsep}{0pt}
\item{Rewarding miners based on number of coins owned rather than amount of electricity and computing resources spent - effectively replacing the operating costs of proof-of-work mining with the capital costs of holding coins;}
\item{Making mining rewards proportionate to number of coins held and time passed since they last generated a reward, making them akin to “interest payments” on the miner's coin holdings.}
\end{enumerate}   
Since there are virtually no operating costs in proof-of-stake, transaction fees can be far lower than for proof-of-work in the long run, regardless of transaction volumes. And because all proof-of-stake miners earn the same rate of return regardless of computing hardware or electricity costs, there is no gradual centralization. 
 
The paper also rebutted the proof-of-work community's various ``nothing at stake'' objections to proof-of-stake and mathematically demonstrated that all commonly cited attack vectors would fail against NeuCoin's design, which increases security relative to Peercoin and other existing proof-of-stake currencies in numerous ways, including:

\begin{enumerate}
\setlength{\itemsep}{0pt}
\item{High mining rewards, lower \textit{minimum stake age} and omission of coin age from the mining equation all incentivize nodes to continuously stake coins over time}
\item{Decreased \textit{\textit{block time}}, improving user experience and enhancing security against some attack vectors}
\item{Causing the \textit{stake modifier} parameter to change over time for each stake, to substantially increase security against precomputation attacks}
\item{Utilizing a client that punishes nodes that attempt to mine on multiple branches with duplicate stakes}
\end{enumerate}
 
As a result, NeuCoin's design solves both the mounting cost and centralization problems of proof-of-work, and the security and centralization problems with earlier proof-of-stake coins. As such, NeuCoin is the first peer-to-peer cryptocurrency, regardless of technology, that is secure, cost-efficient and decentralized in the long run.

\newpage

\bibliographystyle{abbrv}
\bibliography{bibli}

\end{document}